\documentclass[aps,prx,twocolumn,floatfix,superscriptaddress,showpacs,longbibliography]{revtex4-1}
\usepackage{graphicx}
\usepackage{amsmath}
\usepackage{amssymb}
\usepackage{amsfonts}
\usepackage{bm}        % for math
\usepackage{dsfont}
\usepackage[mathscr]{eucal}
\usepackage[colorlinks, linkcolor=OrangeRed,citecolor=RoyalBlue,urlcolor=NavyBlue]{hyperref}
\usepackage[normalem]{ulem}
\usepackage[all]{hypcap}
\usepackage[dvipsnames,usenames,table]{xcolor}

\def \H{\mathcal{H}}

\def\be{\begin{equation}}
\def\ee{\end{equation}}
\def\bea{\begin{eqnarray}}
\def\eea{\end{eqnarray}}

\newcommand{\bk}{{\bf k}}

\newcommand{\bb}{{\bf b}}

\newcommand{\bs}{{\bf s}}

\newcommand{\bp}{{\bf p}}
\newcommand{\br}{{\bf r}}
\newcommand{\bD}{{\bf D}}
\newcommand{\bB}{{\bf B}}

\newcommand{\up}{\uparrow}
\newcommand{\down}{\downarrow}
\newcommand{\diff}{\boldsymbol{\nabla}}

\newcommand{\beq}{\begin{eqnarray}}
\newcommand{\eeq}{\end{eqnarray}}
\newcommand{\beqq}{\begin{eqnarray*}}
\newcommand{\eeqq}{\end{eqnarray*}}

\def\bs#1{\boldsymbol{#1}}

\begin{document}
\title{ Inhomogeneous Weyl and Dirac semimetals: Transport in axial magnetic fields  and Fermi arc surface states from pseudo Landau levels}

\author{Adolfo G. Grushin}
\affiliation{Department of Physics, University of California, Berkeley, CA 94720, USA}
\author{J\"orn W. F. Venderbos}
\affiliation{Department of Physics, Massachusetts Institute of Technology, Cambridge, MA 02139, USA}
\author{Ashvin Vishwanath}
\affiliation{Department of Physics, University of California, Berkeley, CA 94720, USA}
\affiliation{Department of Physics, Harvard University, Cambridge, MA, 02138, USA}
\author{Roni Ilan}
\affiliation{Department of Physics, University of California, Berkeley, CA 94720, USA}
\affiliation{Raymond and Beverly Sackler School of Physics and Astronomy, Tel Aviv University, Tel Aviv 69978, Israel}

\begin{abstract}
Topological Dirac and Weyl semimetals have an energy spectrum that hosts Weyl nodes appearing in pairs of opposite chirality. 
Topological stability is ensured when the nodes are separated in momentum space and unique spectral and transport properties follow.
In this work we study the effect of a space dependent Weyl node separation, which we interpret as an emergent background axial vector potential, on the electromagnetic response and the energy spectrum of Weyl and Dirac semimetals. 
This situation can arise in the solid state either from inhomogeneous strain or non-uniform magnetization and can also be engineered in cold-atomic systems.
Using a semiclassical approach we show that the resulting axial magnetic field $\mathbf{B}_{5}$ is observable through an enhancement of the conductivity as 
$\sigma\sim \mathbf{B}_{5} ^{2}$ due to an underlying chiral pseudo magnetic effect. 
We then use two lattice models to analyze the effect of $\mathbf{B}_5$ on the spectral properties of topological semimetals.
We describe the emergent pseudo-Landau level structure for different spatial profiles of $\mathbf{B}_5$, revealing that 
(i) the celebrated surface states of Weyl semimetals, the Fermi arcs, can be reinterpreted as $n=0$ pseudo-Landau levels resulting from a $\mathbf{B}_5$ confined to the surface
(ii) as a consequence of position-momentum locking a bulk $\mathbf{B}_5$ creates pseudo-Landau levels interpolating in real space between Fermi arcs at opposite surfaces and
(iii) there are equilibrium bound currents proportional to $\mathbf{B}_{5}$ that average to zero over the sample, which are the analogs of bound currents in magnetic materials.
We conclude by discussing how our findings can be probed experimentally.

\end{abstract}\maketitle
\section{Introduction}
Electronic and lattice degrees of freedom are inevitably intertwined in solid state physics~\cite{LANDAU1984xii}. 
With the advent of graphene~\cite{CNGP09} a remarkable effect was soon acknowledged: elastic deformations of the lattice originating from strain can couple to the low-energy Dirac quasiparticles of graphene as a pseudo or \emph{axial} magnetic vector potential~\cite{Vozmediano2010109,JP07,Amorim20161}.
In order to preserve time-reversal symmetry, the axial vector potential couples to the two valleys with an opposite sign. Spatially inhomogeneous strain generates an effective axial-magnetic field $\mathbf{B}_{5}$, and gives rise to a pseudo-Landau level spectrum at low energies~\cite{GKG10}. Strain-induced pseudo-Landau levels have been observed with scanning tunneling microscopy both in real~\cite{LBM10} and artificial graphene~\cite{GKW12,PGF13}.
The realization of Landau levels without real external magnetic fields results in effective fields as high as$~300$T, and is a direct consequence of the Dirac nature of the carriers, exemplifying the unique response of these class of systems to strain.

In this respect, Weyl and Dirac semimetals in three dimensions (3D)~\cite{Turner:2013tf,Hosur2013} are expected to behave as 3D cousins of graphene and host similar effects. A Weyl semimetal is a state with pairs of band touching points with linear dispersion, also called Weyl nodes. The Weyl nodes are sources and sinks of Berry curvature, i.e., Berry curvature monopoles, in momentum space. The charge of a monopole defines the chirality of the corresponding node, and the two partners of a pair of nodes must have opposite chirality. Since monopoles can only annihilate in pairs, Weyl nodes are topologically protected as long as they are separated in momentum space by a vector $\mathbf{b}$. Dirac semimetals can be then regarded as special cases of Weyl semimetals, where nodes of opposite chiralities are located at the same momentum but additional symmetries constrain the system to remain gapless.

The vector $\mathbf{b}$ can be alternatively interpreted as an {\it axial} gauge field since it couples with an opposite sign to Weyl nodes of opposite chirality~\cite{Volovik:1999da,Liu:2013kv,Grushin:2012cb,Zyuzin2012kl,Goswami:2013jp,Ramamurthy:2014uh}. 
This interpretation suggests that if $\mathbf{b}$ is varied in space it will generate a nonzero axial magnetic field $\mathbf{B}_5=\mathbf{\nabla}\times \mathbf{b}$ that couples to fermions of opposite chirality with an opposite sign. This observation, together with the recent theoretical evidence that effective gauge fields can emerge in strained Weyl semimetals~\cite{CFL15,SS16,SIC16,RKY16} are two important motivations of our study.
% the effect of axial magnetic fields on Weyl and Dirac semimetals.

The main purpose of this work is therefore to address the effect of an axial magnetic field $\mathbf{B}_{5}$ arising in any type of Weyl and Dirac semimetal. 
We will first discuss that the physical mechanism for the emergence of $\mathbf{B}_{5}$ depends on the presence or absence of time reversal symmetry.
When a Weyl semimetal preserves time reversal symmetry, each pair of nodes has a time reversed partner (hence the minimal number of nodes is four). In this case, $\mathbf{B}_5$ can be generated by an inhomogeneous strain profile and has an opposite sign for each time reversed pair of nodes such that time reversal symmetry is preserved. In contrast, time reversal breaking Weyl semimetals are not subjected to this constraint and $\mathbf{B}_5$ can emerge either from an inhomogeneous magnetization or strain~\cite{Liu:2013kv,CFL15}. Then, a richer set of phenomena can occur, with the example of a finite angular momentum for the electronic states~\cite{CCG14}.

Here we naturally unify these mechanisms under a single framework by considering any Weyl semimetal in the presence of a spatially varying Weyl node separation. 
We discuss the effect of inhomogeneous nodal separation on 
{\it (i)} the changes in the spectrum due to the emergence of pseudo-Landau levels and 
{\it (ii)} transport, and in particular, the effect on the conductivity.
In exploring {\it (i)}, we will show that the Fermi arcs, the characteristic surface states of topological semimetals~\cite{Wan2011,Burkov2011}, 
can be reinterpreted as the chiral $n=0$ pseudo-Landau level associated with a large $\mathbf{B}_5$ confined to the surface.
On the other hand, the presence of a finite $\mathbf{B}_5$ in the bulk also creates pseudo-Landau levels. The $n=0$ pseudo-Landau level of the bulk is then smoothly connected to the surface arcs. We find that this follows from a known phenomenon in the two dimensional quantum Hall effect, sometimes referred to as position-momentum locking; the average center position of a Landau level wavefunction in one planar direction, say $\left\langle y \right\rangle$, is determined by its momentum in the perpendicular direction, $k_{x}$. Such a concept, considered recently in some detail for real magnetic fields in Weyl semimetals~\cite{BQ16,OK16} will also be of use here. 
We will find that these results significantly depart from the naive expectation gained from studies of strained graphene~\cite{Vozmediano2010109,Amorim20161,JP07} from which the bulk 0-th pseudo-Landau levels is expected to have opposite chirality with respect to the Fermi arcs~\cite{Liu:2013kv,PCF16}.
Regarding {\it (ii)}, we will show, based on both field theoretic arguments and a rigorous semiclassical calculation, 
that a finite bulk $\mathbf{B}_5$ enhances the conductivity of a Weyl semimetal as $\sigma\sim \mathbf{B}_5^2$. 
This response is related to a charge anomaly much like the topological negative magnetoresistance $\sigma\sim \mathbf{B}^2$ is rooted in the chiral anomaly~\cite{NielNino83}. 

Moreover, we find that a time reversal breaking Weyl semimetal may support bound currents originating in a chiral pseudo magnetic effect driven by $\mathbf{B}_5$. This effect exists in equilibrium, unlike the standard chiral magnetic effect~\cite{FKW08,Kharzeev2014,Zyuzin:2012ca,Grushin:2012cb,Zyuzin2012kl,Goswami:2013jp,ZJN13,Landsteiner:2014fw} which is strictly zero in a non-dynamical situation~\cite{Vazifeh:2013fe,Chang2015,ZMS16,Ma2015,Zub16}.

The advances made on the experimental front in recent years allows us to explore the feasibility and detectability of these effects while considering a wide range of possible platforms. While Dirac and Weyl semimetals material growth thrives in condensed matter~\cite{KimiKim2013,LKZ16,Huang2015b,ZhangXu2015,XKL2015,YangLi2015,
SAW2015,Du2015,ZGC2015,LiHe2015,Weng2015,
Huang2015,SIN15,Lv2015,XAB15,LvXu2015,YLS15,
Xu2013,Liu2014a,SJB15,Xu2015,Neupane2014,Borisenko2014,Yi2014,Liu2014b,Liang2015,He2014,Feng2014,BEG15,WVK16}, they could also be engineered with cold atoms~\cite{DKLKSB2015}. This provides access to both inversion breaking semimetals that are currently the standard in the solid state (with notable potential exceptions~\cite{BEG15,WVK16}), as well as to time reversal breaking semimetals in cold atoms. We end this paper with an estimate of the magnitude of these effects based on real material parameters and discuss how the phenomena described here might be detected.  

The structure of this paper is as follows.
In section~\ref{sec:effmod} we review how inhomogeneous strain (magnetization) in time-reversal invariant (broken) Weyl and Dirac semimetals
leads to an effective axial magnetic field, or alternatively, a space dependent Weyl node separation.
In section \ref{sec:transport} we predict the enhancement of the 
conductivity of Weyl and Dirac semimetals due to a uniform axial magnetic field, and relate it to the underlying associated chiral pseudo-magnetic effect. 
The latter leads to bound currents flowing within the material and along the boundary, even in equilibrium, as well as an enhanced bulk longitudinal conductivity. 
In section~\ref{sec:spectral} we introduce two lattice models that illustrate the spectral breakdown into pseudo-Landau levels, and explicitly calculate the bound current distribution within a finite sample. 
We describe position momentum locking and show that any equilibrium bulk current is compensated by surface currents. 
As an example of the relevance of our findings to different contexts, 
we discuss the implications of our results on the situation where two Weyl semimetals with different Weyl node separation are brought into proximity.  
Finally, in section \ref{sec:discuss} we provide a discussion and proposals to experimentally observe the discussed phenomena.

\section{Emergence of space dependent node separation\label{sec:effmod}}

In this section we review how a non-uniform magnetization or strain can lead to a space dependent node separation. This discussion of the physical origin of axial magnetic fields in solid states systems will set the stage for our general study presented in the next sections, by introducing the main concepts and notation.

\subsection{Time-reversal symmetric Weyl semimetal}

The existence of Weyl nodes relies on breaking either time-reversal ($\mathcal{T}$) or inversion symmetry ($\mathcal{I}$). If both are present, all energy bands are manifestly twofold degenerate, and, in particular, point nodes must have a degenerate partner with opposite chirality. If $\mathcal{T}$ is broken, the two nodes of a pair are allowed to be separated in momentum space, with the separation determined by $\bb$. If $\mathcal{I}$ is broken but $\mathcal{T}$ is preserved, each pair must have a time-reversed partner, such that the minimum number of Weyl nodes in $\mathcal{I}$-broken Weyl semimetals is four. In an $\mathcal{I}$-broken Weyl semimetal, the Hamiltonian of the low-energy Weyl fermions for each pair is given by
\be 
\H_{\mathrm{Weyl}}=\sum_{i=x,y,z} \sum_{s=\pm} s v_{i} (  k_{i} + s b^i ) \sigma^{i},  \label{eq:weyl}
\ee
where $v_i$ are the Fermi velocities, $\sigma^{i}$ are  Pauli matrices spanning the orbital space of the Weyl fermions, and $s=\pm$ denotes the chirality of the nodes. It is clear from Eq.~\eqref{eq:weyl} that the vector $\bb = (b^x,b^y,b^z)$ quantifies the separation of the two nodes of a pair in momentum space.

In general, in a crystalline material a uniform strain of strength $g$ alters the overlaps between wavefunctions in different orbitals~\cite{CFL15,RKY16}. In Ref.~\onlinecite{RKY16}, for instance, which considered HgTe with spin-orbit induced band inversion in the presence of strain, found that at low energies the spectrum consists of four pairs of Weyl nodes. Due to the presence of strain, the Hamiltonian of each pair then takes the form
\be
\label{eq:Weyl1}
\H_{\mathrm{Weyl}}= \sum_{i,s=\pm}v_{i}(g)\Big[s k_{i}\sigma^{i}+b^{i}(g)\sigma^{i}\Big],
\ee
From \eqref{eq:Weyl1}, we observe the effect of strain is twofold. Firstly it modifies the Fermi velocities through $v_{i}(g)\sim \sqrt{g}$. Secondly, it controls the Weyl node separation through the vector $\mathbf{b}$ with magnitude $|\mathbf{b}(g)|\sim \sqrt{g}$. Both of these effects are analogous to those predicted in graphene~\cite{JSV12, Vozmediano2010109, Amorim20161}.

Ref.~\cite{RKY16} analyzed the effect of strain as an homogeneous perturbation characterized by $g$. However in practice, strain can have a non uniform profile $g(\mathbf{r})$. This can occur, for instance, when there is a lattice mismatch between a substrate and the semimetal sample or alternatively via chemically induced strain. In the former scenario, for thin enough samples, it can be expected that the strain profile relaxes smoothly as a function of the distance from the substrate (quantitative estimates will be provided in section \ref{sec:discuss}). Let us then choose the direction of the strain inhomogeneity to be $y$ such that $g(\mathbf{r})=g(y)$, and assume that $g(y)$ changes slowly with $y$, on a length scale that is much smaller than any other microscopic length scale.  In such a perturbative regime, $g(y) \simeq g_{0}+\delta g(y)$ with $\delta g (y) /g_{0} \ll1$, it is reasonable to expand  Eq.~\eqref{eq:Weyl1} to lowest order in $\delta{g}(y)/g_{0}$. Using that $v[g(y)] \simeq v(g_{0})(1 + \delta g (y)/2g_0)$ and $b^{i}[g(y)] \simeq b_{i}(g_0)(1+\delta g(y)/2g_0)$, one
obtains
\be
\label{eq:Weyl2}
\H_{\mathrm{Weyl}}\simeq \sum_{i,s=\pm} v(g_0)\sigma^{i}\left[s i\partial_{i}+b^{i}(g_0)\left(1+\dfrac{\delta g(y)}{g_{0}}\right)\right].
\ee
We first note that the effect of the space dependent Fermi velocity due to strain~\cite{JSV12} is of order $\frac{\delta g(y)}{g_0} \partial_{i}$, so it can be neglected in the low-energy linear approximation. An inhomogeneous velocity has been predicted recently to result in an anomalous Hall signal transverse to the direction in which the strain is applied~\cite{KKL15,YPZ15}. However, these works neglected the effect of emergent inhomogeneous Weyl node separation that, according to the above analysis, it is of lower order in $\delta g(x)$ and therefore dominant. 

Motivated by the above effective model, we consider the following real space Hamiltonian
\be
\label{eq:Weyleff}
\H_{\mathrm{Weyl}}\simeq \sum_{i,s={\pm}}v\Big[ si\partial_{i}-b^{(0)}_i+\delta b_{i}(\mathbf{r})\Big] \sigma^{i},
\ee
where $\mathbf{b}(\mathbf{r})\equiv \mathbf{b}^{(0)}-\delta \mathbf{b}(\mathbf{r})$ encodes the strain profile.
From such a model with a single pair of nodes, the Dirac semimetal and the inversion breaking Weyl semimetal can be constructed by adding time-reversal partners of Eq.~\eqref{eq:Weyleff} in the appropriate location in momentum space in order to restore time reversal symmetry.

We note that in the literature, the vector $\mathbf{b}$ has been referred to, depending on the context, as 
chiral, axial or pseudo vector potential. In this work, we will refer to $\mathbf{b}$ as the axial vector potential as custom in high-energy physics where this field was first discussed.

\subsection{Time-reversal breaking Weyl semimetal}

In $\mathcal{T}$-breaking Weyl semimetals, the axial vector potential $\mathbf{b}$ that separates the Weyl nodes in momentum space can have its physical origin in a finite magnetization. A hallmark example of  a Weyl semimetal of this kind is the proposal by Burkov and Balents~\cite{Burkov:2011de}, which is based on a topological insulator-trivial magnetic insulator heterostructures. Another example is the Weyl semimetal state induced by magnetic order at a quadratic band crossing point on pyrochlore iridates \cite{SMB14}. In addition, a time-reversal breaking Weyl semimetal has been argued to be consistent with ARPES measurements on YbMnBi$_2$~\cite{BEG15}. Other promising recent proposals~\cite{WVK16} suggest, relying on \emph{ab-initio} calculations, that magnetic Heusler compounds can host Weyl quasiparticles. Remarkably, alloys of the latter kind could realize the minimal model of a $\mathcal{T}$-breaking Weyl semimetal with only two nodes, and a node separation that spans a significant fraction of the Brillouin zone.

Similar to other magnetic materials, the magnetization that induces the Weyl semimetal state by separating two nodes of opposite chirality in momentum space, may itself be non-uniform and become a function of position~\cite{HZJ14}. As a result, the Weyl node separation becomes a function of position, generating an axial magnetic field strength. To lowest order in the spatial variation of the Weyl node separation (i.e., magnetization), the effective model~\eqref{eq:Weyleff} will describe this state. 
We note as well that interfaces between magnetic domains, either abrupt or continuous could
be modeled by the spatial profiles of $\mathbf{b}$ that we consider in this work. In addition, it is plausible to expect that coating one of the surfaces with a ferromagnetic layer or enough surface magnetic dopants will also result in a spatially varying magnetization. The successful realization of magnetically doped topological insulators~\cite{SBK13} suggests that similar techniques could be applied in Weyl samples as well.

In addition to an inhomogeneous magnetization, a non-uniform strain can generate an axial magnetic field in $\mathcal{T}$-breaking Weyl semimetals. Once $\mathcal{T}$ is broken and a finite $\mathbf{b}$ exists, strain can allow to effectively change $\mathbf{b}$ from its bare value, thus changing the separation of Weyl nodes in momentum space. Note, however, that strain on its own cannot break time-reversal symmetry and thus cannot induce a finite $\mathbf{b}$ from  $\mathbf{b}=0$.

\subsection{Effective realization in cold atomic systems }
While the above sections discuss realizations of inhomogeneous Dirac and Weyl semimetals in the solid state,
our discussion is also relevant for cold atomic systems~\cite{DKLKSB2015}.
Two appealing aspects of the latter is that different topological properties can be extracted easily from experimental data~\cite{DG13,ALS14,JMD14} and models that respect or break time reversal could be engineered.

There are two further advantages of considering cold atomic systems that are relevant to probe the effects we discuss.
Firstly, engineering a space dependent Weyl node separation is experimentally feasible. 
A natural way to achieve this is to smoothen the confining trapping potential that contains the lattice~\cite{GDD13}.
Although plausible, it does not offer the degree of control that will expose all of the salient features we are after, but only a subset of them.
A more controlled and realistic approach is to create domain walls between topological states~\cite{VGB15} as already
described in Ref.~\cite{GJM16} for Chern insulators.
%
%Since Weyl semimetals are themselves a collection of Chern insulator planes, this is likely to be a route to realize the physics we propose.
%
We elaborate on more particulars of this proposal in the concluding section.

The second advantage of cold-atomic systems is the controlled experimental access to transport properties that was recently discussed in the context of Chern insulators~\cite{PZO16}\footnote{We thank N. Goldman for pointing out the relevance of this protocol to our proposal}.
The key observable is the center-of-mass velocity $\mathbf{v}_{\mathrm{c.m.}}$ since it is simple to measure experimentally 
and is connected to the current density $\mathbf{j}$ and particle density $n$ through~\cite{PZO16}
\be
\label{eq:vcm}
\mathbf{v}_{\mathrm{c.m.}}=\dfrac{\mathbf{j}}{n}.
\ee
Although in general $n$ also depends on external fields, in our particular case it acts
as an unimportant constant factor as we describe below.
Thus, via Eq.~\eqref{eq:vcm} it is possible to directly probe the longitudinal conductivity. 

\section{Enhanced topological longitudinal conductance \label{sec:transport}}

Many peculiar transport properties have been attributed to topological semimetals~\cite{Turner:2013tf,Hosur2013}, while not many of them have proven easy to measure. 
One particularly striking feature that has been confirmed in several experiments
~\cite{KimiKim2013,LKZ16,Huang2015b,ZhangXu2015,XKL2015,YangLi2015,
SAW2015,Du2015,ZGC2015,LWL15,LiHe2015,Weng2015,
Huang2015,SIN15,Lv2015,XAB15,LvXu2015,YLS15,
Xu2013,Liu2014a,SJB15,Xu2015,Neupane2014,Borisenko2014,Yi2014,Liu2014b,Liang2015,He2014,Feng2014} 
is the appearance of a non-saturating negative magnetoresistance~\cite{Son:2013jz,Burkov:2015wt}. 
Conductance is one of the most accessible experimental observables to characterize the properties of a material.
Therefore, we aim here to explore the effect of an inhomogeneous Weyl node separation on this quantity.
We show that a space dependent  axial vector potential $\mathbf{b}$ 
enhances the longitudinal conductance of a Weyl semimetal through its corresponding axial magnetic field $\mathbf{B}_{5} = \boldsymbol\nabla\times\mathbf{b}$.
We begin by introducing a coupling to an external electromagnetic field and discussing the role of gauge invariance.
We follow this by extending a quantum field theory argument previously used in Ref.~\cite{LKZ16} to include pseudo gauge fields. We then support
our claims with a more rigorous semiclassical calculation by solving the semiclassical kinetic equation.

\subsection{Coupling to an external electromagnetic field and the role of gauge invariance}
In order to calculate the conductivity we couple the Hamiltonian~\eqref{eq:Weyleff} to an external electromagnetic field $A_{\mu}=(A_{0},\mathbf{A})$
using the minimal substitution
\be
\label{eq:WeyleffA}
\H_{\mathrm{Weyl}}=\sum_{s={\pm}}v\Big\{ s[i\partial_{i}-eA_{i}(x)]-b_{i}(x) \Big\} \sigma_{i}+A_0,
\ee
The first important property of this Hamiltonian is that left handed ($s=1$) and right handed ($s=-1$) Weyl fermions
are decoupled, 
experiencing a chirality dependent gauge field, $a^{s}_{\mu}=A_{\mu}+s b_{\mu}$, where $b_{\mu}=(0,\mathbf{b}(\mathbf{r}))$.
For what follows it is useful to define the left and right handed field strengths $f^{\mu\nu}_{s}=\partial_{\mu}a^s_{\nu}-\partial_{\nu}a^s_{\mu}$.

The continuity equation for the current for each species $j^{\mu}_s=(n_s,\mathbf{j}_{s})$ is determined by~\cite{bertlmann2000anomalies,Liu:2013kv}
\begin{eqnarray}
\label{eq:contLR}
\partial_{\mu}j^{\mu}_{s} &=& s\frac{e^3}{4\,\pi^2\,\hbar^2} \mathbf{E}_{s} \cdot \mathbf{B}_{s} - \frac{e}{\tau}(n-n_{s}),
\end{eqnarray}
which we now discuss in some depth.
The right hand side of Eq.~\eqref{eq:contLR} is composed by the left and right effective 
electromagnetic fields $\mathbf{E}_{s}=\mathbf{E}+s\mathbf{E}_{5}$ and $\mathbf{B}_{s}=\mathbf{B}+s\mathbf{B}_{5}$,
where $(\mathbf{E},\mathbf{B})$, $(\mathbf{E}_5,\mathbf{B}_5)$, and $(\mathbf{E}_{s},\mathbf{B}_{s})$ are
built out of the gauge field strength $F^{\mu\nu}=\partial_{\mu}A_{\nu}-\partial_{\nu}A_{\mu}$, the axial gauge field strength $f^{\mu\nu}_{b}=\partial_{\mu}b_{\nu}-\partial_{\nu}b_{\mu}$, and $f^{\mu\nu}_{s}$, respectively.
The second term on the right hand side is a scattering term that acts to equilibrate the number density $n_{s}$ of the left and right chiralities~\cite{Parameswaran2014,Behrends:2015ux}
with a typical scattering time $\tau$ \footnote{The origin of this scattering term, and in particular the microscopic origin of $\tau$ is a subtle issue 
due to the apparent non-gauge invariance of the bulk theory. A full field-theoretical description of this term is out of the scope of this work.}. 
The conservation laws for the current $j^{\mu}=\sum_{s}j^{\mu}_{s}$ and axial current  $j_{5}^{\mu}=\sum_{s}s j^{\mu}_{s}$ can be obtained from 
the addition or subtraction of the two equations composing Eq.~\eqref{eq:contLR}.
A first inspection reveals a seemingly striking feature; neither of the two currents is conserved.
The non-conservation of $j^{\mu}_{5}$ is not forbidden, and it is referred to as the chiral anomaly~\cite{bertlmann2000anomalies,TV13,Hosur2013}.
If both $\mathbf{E}_{5}=0$ and $\mathbf{B}_{5}=0$ then one recovers the celebrated chiral anomaly that schematically reads $\partial_{\mu} j^{\mu}_{5}\sim \mathbf{E}\cdot\mathbf{B}$~\cite{NielNino83,bertlmann2000anomalies}.
In this case the vector current satisfies $\partial_{\mu} j^{\mu} = 0$ and thus charge is conserved.
However, it seems that charge is not conserved when either $\mathbf{E}_{5}\neq 0$ or $\mathbf{B}_{5}\neq 0$.
Indeed, from Eq.~\eqref{eq:contLR} we find that the vector current satisfies $\partial_{\mu} j^{\mu} \sim \mathbf{E}_{5}\cdot \mathbf{B} + \mathbf{E}\cdot \mathbf{B}_5$ \cite{Liu:2013kv,Ramamurthy:2014uh}.

In fact, there is no contradiction with current conservation, which can be understood in two different ways.
Firstly, the axial vector potential $\mathbf{b}$ is an observable and thus must be single valued, with a zero vacuum expectation value~\cite{Hosur2013}.
Thus the total flux of the resulting $\mathbf{B}_5$ must be zero over a surface enclosing the entire sample; regions with $\mathbf{B}_{5}$ and $-\mathbf{B}_{5}$ 
compensate each other by generating an equal number of left and right handed fermions separated in real space.
Then, upon applying an electric field charge flows from one region to the other, respecting global charge conservation. 

Second, at the quantum field theory level, the non-conservation of charge is fixed by defining a consistent current that preserves gauge invariance.
The procedure has been described both in the context of high energy physics~\cite{BZ84} and Weyl semimetals~\cite{Landsteiner:2014fw}:
the current calculated above (the covariant current) is complemented by Chern Simons currents (the Bardeen-Zumino polynomials)
of the form $ -\frac{1}{4\pi^2}\epsilon_{\mu\nu\rho\sigma} b^{\nu} F^{\mu\nu}$, which exactly cancel the anomaly and define the consistent current.
This procedure effectively imposes a boundary condition for the spectral flow at the energy of the cut-off that bounds the field theory.
In a lattice system, such a cut-off is a natural quantity, and the spectral flow is bounded by construction.
Note also that this procedure applies wherever there is a finite axial magnetic field $\mathbf{B}_5$ and therefore includes the boundaries
where $\mathbf{b}$ jumps from zero to a finite value or viceversa.

\subsection{Quantum field theory approach}
%--------
%
Our aim is to use the chiral anomaly Eq.~\eqref{eq:contLR}  to find the longitudinal conductance in the presence of a finite axial magnetic field
$\mathbf{B}_5$.
To promote the derivation in Ref.~\cite{LKZ16} we first use that for three-dimensional Weyl fermions $n_{s} = \mu_{s}^3/(6\,\pi^2\,\hbar^3\,v^3)$,
which we employ to rewrite the steady state form of Eq.~\eqref{eq:contLR} as
\begin{equation}
 \mu_{s} = \left[\mu^3-\dfrac{3}{2} s \hbar\,v^3\,e^2\,\tau \,\mathbf{E}_{s} \cdot \mathbf{B}_{s}\right]^{1/3}.
 \label{eq:mu_ca}
\end{equation}
%
%%
%\begin{equation}
% \mu_{s} = \left[\mu^3-s\frac{3}{2} \hbar\,v^3\,e^2\,\tau \,\mathbf{E}_{s} \cdot \mathbf{B}_{s}\right]^{1/3}.
% \label{eq:mu_ca}
%\end{equation}
%%
We now recall that the component of the current parallel to $\mathbf{B}$ and $\mathbf{B}_{5}$ is given by

\begin{equation}
\mathbf{j}= \dfrac{e^{2}}{2\pi^2\hbar^2}\left[\mu_5\mathbf{B}+\mu\mathbf{B}_{5}\right].
 \label{eq:curr}
\end{equation}
The first term is the chiral magnetic effect~\cite{FKW08,Kharzeev2014} and has been thoroughly studied in the context of Weyl semimetals~\cite{Zyuzin:2012ca,Grushin:2012cb,Zyuzin2012kl,Goswami:2013jp,ZJN13,Landsteiner:2014fw}.
This term must be zero in equilibrium~\cite{Vazifeh2013,Goswami:wl,Chen:2013ep,Ma2015,ZMS16} since it is proportional to a chemical potential
imbalance $\mu_{5}=\frac{1}{2}\sum_{s}s\mu_s$.
The second term on the left hand side is the key to our results:
it represents the analog of the chiral anomaly in the presence of an axial magnetic field $\mathbf{B}_{5}$ and can be finite in equilibrium.
In the context of dense relativistic matter it was shown that it is possible to generate an axial current $\mathbf{j}_{5}$ from a vector field $\mathbf{B}$ with a conductivity proportional
to the chemical potential $\mu=\frac{1}{2}\sum_s\mu_s$~\cite{SZ04}.
It follows that there is a contribution to the vector current $\mathbf{j}$ generated by an axial field with the \emph{same} coefficient~\cite{Landsteiner2013}.
We refer to this term as the chiral pseudo-magnetic effect.

In order to obtain the longitudinal conductivity we combine Eqs.~\eqref{eq:curr} and \eqref{eq:mu_ca},
\begin{eqnarray}
 \mathbf{j}&=&\dfrac{e^{2}}{h^2}\sum_{s}s\mu_s\mathbf{B}_{s}\\ 
 &=&\dfrac{e^{2}}{h^2}\sum_{s}s\mu\left[1-s\frac{3}{2\mu^{3}} \hbar\,v^3\,e^2\,\tau \,\mathbf{E}_{s} \cdot \mathbf{B}_{s}\right]^{1/3}\mathbf{B}_{s}.
 \label{eq:dens}
\end{eqnarray}
Assuming that 
\begin{eqnarray}
\frac{3}{2\mu^3} \hbar\,v^3\,e^2\,\tau \,\mathbf{E}_{s} \cdot \mathbf{B}_{s}\ll1,
\end{eqnarray}
and for the case where $\mathbf{B}=B\hat{z}$ and $\mathbf{B}_5=B_5\hat{z}$, $\mathbf{B}_{s}=(B+sB_{5})\hat{z}$ and $E_{5}=0$, we obtain
\begin{eqnarray}
\label{eq:NMR}
 j_{z}&=&\frac{\,v^3\,e^4\,\tau}{4\pi^{2}\mu^{2}\hbar} \,E (B^2+B_{5}^{2})+\dfrac{e^{2}}{2\pi^2\hbar^2}\mu B_{5},
 \label{eq:dens}
\end{eqnarray}
to linear order in $\mathbf{E}$. 
The first term is a transport (or free) current.
For the case when $B_{5}=0$ it reproduces the chiral anomaly enhanced magneto-conductivity~\cite{Son:2013jz}.
The central result here is that for $\mathbf{B}_{5}\neq 0$ there is a contribution to the bulk longitudinal conductivity
coming entirely from the spatial dependence of $\mathbf{b}(\mathbf{r})$:
inhomogeneous strain or magnetization enhances conductance in a Weyl or a Dirac semimetal.

We now provide a physical interpretation of the second term as a magnetization (or bound) current.
Recall that $\mathbf{b}(\mathbf{r})$ is analogous to a finite magnetization $\mathbf{M}(\mathbf{r})$.
Thus local bound currents given by $\mathbf{j}_{b}\propto\bs{\nabla}\times \mathbf{M}(\mathbf{r}) \propto \mathbf{B}_5$ are allowed, consistent with 
the form of the second term in Eq.~\eqref{eq:NMR}.
Note that in the related context of the quantum Hall effect, a finite bound current proportional to $\mu$ and the curl of the magnetization is expected on general grounds~\cite{CHR96}.
In that case, the bound current has its origin on the edge states, while here they can be associated with the bulk.

We remark that the second term in Eq.~\eqref{eq:NMR} highlights an important difference between semimetals that respect time reversal symmetry
and those that do not.
In the former the sum over time reversed pairs of nodes will cancel out the $\mu \mathbf{B}_{5}$ term.
In Section~\ref{sec:Boltzmann} we provide further arguments that support the interpretation of this term as a bound current, 
while in Section~\ref{sec:spectral} we corroborate these findings by numerically studying the bound current profile within specific lattice models.

\subsection{Boltzmann equation approach\label{sec:Boltzmann}}

In this section we outline a different and more rigorous derivation of Eq.~\eqref{eq:NMR} that relies on solving the Boltzmann equation. 
Within this approach, the semiclassical equations of motion are extended to include an anomalous velocity term that arises due to the existence of a non-zero Berry curvature~\cite{Xiao2010}. 
Typically, for a Weyl semimetal, the equations of motion are written for a single flavor of chiral fermions accounting for the physics in momentum space centered around a particular Weyl node~\cite{Son:2013jz}.
Both chiralities feel the same electric and magnetic fields.
The key difference in the present analysis is that the effective external fields are now chirality dependent due to the axial vector potential; both chiralities are still decoupled but feel different effective fields.

More precisely, the effect of the effective magnetic fields $\mathbf{B}_{s}=(\mathbf{B}+s\mathbf{B}_{5})$ on the left ($s=+1$) and right ($s=-1$) chiralities can be incorporated by promoting
the semiclassical equations of motion~\cite{Xiao2010} to
\bea
\label{eq:rdot}
&&\dot{\mathbf{r}}_{s}=\partial_{\mathbf{p}}\mathcal{E}_{\mathbf{p}}^{s}+\dot{\mathbf{p}}\times \boldsymbol{\Omega}^{s}_{\bf{p}}, \\
\label{eq:pdot}
&&\dot{\mathbf{p}}_{s}=e\mathbf{E}+\frac{e}{c}\dot{\mathbf{r}}\times \mathbf{B}_{s}.
\eea
Here $\mathcal{E}^{s}_{\mathbf{p}}=\varepsilon_{\mathbf{p}}^{s}-\mathbf{m}^{s}_{\mathbf{p}}\cdot\mathbf{B}_{s}$ and $\boldsymbol{\Omega}^{s}_{\bf{p}}$ are, respectively, the dispersion relation, which includes a correction
due to the magnetic orbital moment $\mathbf{m}^s_{\mathbf{p}}$, and the corresponding Berry curvature for each chirality $s$.
For each chirality the unperturbed dispersion relation of the upper band is $\varepsilon_{\mathbf{p}}^{s}= sv|\mathbf{p}|$.
The Berry curvature and magnetic orbital moment in this case take the simple form $\boldsymbol{\Omega_{\mathbf{p}}}^{s}=s\frac{1}{2|\mathbf{p}|^{2}}\mathbf{\hat{p}}$ and $\mathbf{m}^{s}_{\mathbf{p}}=-e v |\mathbf{p}| \boldsymbol{\Omega_{\mathbf{p}}}^{s}$.
We emphasize that, consistent with the discussion in Sec.~\ref{sec:effmod}, we neglect effects of the higher order corrections due to the inhomogeneous Fermi velocity 
which can enter through a space dependent Berry curvature~\cite{SN98,KKL15,YPZ15}.
The distribution function for each chirality  $f^{s}_\mathbf{p}$ 
satisfies a semiclassical kinetic equation
\be
\frac{\partial f^{s}_{\mathbf{p}}}{\partial t}+\dot{\bf{r}}\cdot\frac{\partial f^{s}_{\mathbf{p}}}{\partial \bf{r}}+\dot{\bf{p}}\cdot\frac{\partial f^{s}_{\mathbf{p}}}{\partial \bf{p}}=I_{coll}\{f \},
\ee
where $I_{coll}\{f\}$ is the collision integral.
Using Eqs.~\eqref{eq:rdot} and \eqref{eq:pdot} we can write 
\be
\dot{\mathbf{r}}_{s}=D_{\mathbf{p},s}^{-1}\left(\mathbf{v}^{s}_{\mathbf{p}}+e\mathbf{E}\times \mathbf{\Omega}^{s}_{\mathbf{p}}+\frac{e}{c}(\mathbf{\Omega}^{s}_{\mathbf{p}}\cdot \mathbf{v}^{s}_{\mathbf{p}})\mathbf{B}^{s}\right),
\ee 
where $D_{\mathbf{p},s}=(1+\frac{e}{c}\mathbf{B}^{s}\cdot{\boldsymbol{\Omega}}^{s}_{\bf{p}})$ and $\mathbf{v}^{s}_{\bf{p}}=\partial \mathcal{E}^{s}_{\mathbf{p}}/\partial \mathbf{p}$ is the perturbed velocity. 
We find that the contribution of a single chirality to the current density is given by
\bea
\nonumber
\mathbf{j}^{s}&=&e\int \dfrac{d^3 p}{(2\pi)^3}  D_{\mathbf{p},s} \mathbf{\dot{r}}_{s}f^{s}_\mathbf{p}(\mathcal{E}^s_{\mathbf{p}}),\\ 
\nonumber
&=&e\int \dfrac{d^3 p}{(2\pi)^3} \left(\mathbf{v}^{s}_{\mathbf{p}}+e\mathbf{E}\times \mathbf{\Omega}^{s}_{\mathbf{p}}+\frac{e}{c}(\mathbf{\Omega}^{s}_{\mathbf{p}}\cdot \mathbf{v}^{s}_{\mathbf{p}})\mathbf{B}^{s}\right)f^s_\mathbf{p}(\mathcal{E}^s_{\mathbf{p}}).\\
\label{eq:currentk}
\eea
Within the relaxation time approximation $I_{coll}\{f\}=-(f^{s}_\mathbf{p}-f^{0}_{\mathbf{p},s})/\tau$ where $\tau$ is a scattering time~\footnote{In general, $\tau=\tau_{\mathrm{inter}}^{-1}+\tau_{\mathrm{intra}}^{-1}$. Since for realistic situations $\tau_{\mathrm{inter}}\gg\tau_{\mathrm{intra}}$~\cite{Behrends:2015ux}  we take $\tau\sim\tau_{\mathrm{intra}}$}  and  $f^{0}_{\mathbf{p},s}(\mathcal{E}_{\mathbf{k},s})$ is the equilibrium distribution function to be evaluated at the modified dispersion relation $\mathcal{E}_{\mathbf{k},s}$.  
We are interested in a stationary and homogeneous solution to the kinetic equation, which takes the form 
\be
\dot{\bf{p}}\cdot\frac{\partial f^{s}_{\mathbf{p}}}{\partial \bf{p}}=-\frac{(f^s_\mathbf{p}-f^{0}_{\mathbf{p},s})}{\tau}.
\ee
Expanding the left hand side of the previous equation to lowest order in the fields and rearranging terms, we obtain
\be
\label{eq:equilibrium1}
f^s_\mathbf{p}=f_{\mathbf{p},s}^0-\tau D_{\mathbf{p},s}^{-1}\left(e\mathbf{E}+\frac{e}{c}\mathbf{v}^{s}_{\mathbf{p}}\times\mathbf{B}^{s}+\frac{e^2}{c}(\mathbf{E}\cdot\mathbf{B}^{s})\mathbf{\Omega}^s_{\mathbf{p}}\right) \frac{\partial f^{0}_{\mathbf{p},s}}{\partial \mathbf{p}}.
\ee
Inserting the first term in Eq.~\eqref{eq:equilibrium1} into Eq.~\eqref{eq:currentk} and ignoring the contribution transverse to electric field leads to
\bea
\mathbf{j}^s_0&=&e\int \dfrac{d^3 p}{(2\pi)^3} \frac{e}{c}(\mathbf{\Omega^{s}_{\mathbf{p}}}\cdot \mathbf{v}^{s}_{\mathbf{p}})\mathbf{B}^{s}f^0_{\mathbf{p},s}(\mathcal{E}^s_{\mathbf{p}}).
\eea
To leading order in the magnetic field, we can make the replacement $\mathcal{E}^{s}_{\mathbf{p}} \to \varepsilon^s_{\mathbf{p}}$ in both the equilibrium distribution function and the definition of the
velocity. 
The integral results in

\bea
\label{eq:CPME}
\mathbf{j}^{s}_0&=& s\frac{e^2}{h} \mu\mathbf{B}^{s},
%s\frac{e^2}{4\pi^2\hbar^2}\mu\mathbf{B}^{s}
\eea
with $\mu$ the chemical potential, consistent with earlier findings~\cite{ZJN13,SS16}.
Summing Eq.~\eqref{eq:CPME} over chiralities results in the second term in Eq.~\eqref{eq:NMR}.

Remarkably, we have also performed a rate of entropy production calculation in the spirit of Ref.~\cite{Son:2013jz} and find this term to be absent.
The reason for this absence is that the rate of entropy production is determined exclusively by the \textit{free} current density rather than the total current density.
This result implies, as anticipated above, that Eq.~\eqref{eq:CPME} should be physically interpreted as a bound current proportional to the curl of the magnetization in the system and as such 
does not contribute to transport.

Next, we insert the second term in Eq.~\eqref{eq:equilibrium1} into Eq.~\eqref{eq:currentk}
to calculate the correction to the current corresponding to the deviation $\delta f^s_{\mathbf{p}}= f^s_\mathbf{p}-f^{0}_{\mathbf{p},s}$ up to first order
in $\mathbf{E}$ and second order in $\mathbf{B}_{s}$, which reads
\bea
\label{eq:currentsemi}
\nonumber
\delta\mathbf{j}^s&=&-e\tau\int\dfrac{d^3 p}{(2\pi)^3}D_{\mathbf{p},s}^{-1}\left[e\mathbf{v}^{s}_{\mathbf{p}}\left(\mathbf{E}\cdot\dfrac{\partial f^s_\mathbf{p}}{\partial\mathbf{p}}\right)\right.+\\
\nonumber
&+&\dfrac{e^2}{c}\mathbf{v}^{s}_{\mathbf{p}}\left(\mathbf{E}\cdot\mathbf{B}^{s}\right)\left(\boldsymbol{\Omega}^{s}_{\bf{p}}\cdot\dfrac{\partial f^s_\mathbf{p}}{\partial\mathbf{p}}\right)+\\
\nonumber
&+&\dfrac{e^2}{c}\mathbf{B}^s\left(\boldsymbol{\Omega}^{s}_{\bf{p}}\cdot\mathbf{v}^{s}_{\mathbf{p}}\right)\left(\mathbf{E}\cdot\dfrac{\partial f^s_\mathbf{p}}{\partial\mathbf{p}}\right)+\\
&+&\left.\dfrac{e^3}{c^2}\mathbf{B}^s\left(\boldsymbol{\Omega}^{s}_{\bf{p}}\cdot\mathbf{v}^{s}_{\mathbf{p}}\right)\left(\mathbf{E}\cdot\mathbf{B}^s\right)\left(\boldsymbol{\Omega}^{s}_{\bf{p}}\cdot\dfrac{\partial f^s_\mathbf{p}}{\partial\mathbf{p}}\right)\right].
\eea
Using that for the lower band $\boldsymbol{\Omega_{\mathbf{p}}}^{s}=-s\frac{1}{2|\mathbf{p}|^{2}}\mathbf{\hat{p}}$ and $\mathbf{m}^{s}_{\mathbf{p}}=-e v |\mathbf{p}| \boldsymbol{\Omega_{\mathbf{p}}}^{s}$
and after tedious but straightforward manipulations we obtain
\be
\label{eq:equilibrium}
\delta j^s_z = -\tau\left( \dfrac{1}{3}e^2v^2\nu(\mu)+\dfrac{v^3 e^4}{30\pi^2 \hbar \mu^2} B_{s}^{2}\right)E,
\ee
where $\nu(\varepsilon)=\varepsilon^2/2\pi^2\hbar^3v^3$ is the unperturbed density of states of a Weyl node at energy $\varepsilon$.
The first term encodes the conductivity due to a finite Fermi surface~\cite{BHB11,DHM15}.
The second term is novel to this work and upon summing chiralities leads to a longitudinal contribution to the conductivity that reads
\be
\label{eq:equilibrium2}
\sigma_{zz} = \tau\dfrac{v^3 e^4}{15\pi^2 \hbar \mu^2} B^2_{5},
\ee
for $B=0$ and represents the main result of this section.
Up to the numerical coefficient, this contribution has exactly the same form as the first term of Eq.~\eqref{eq:NMR} confirming that an inhomogeneous strain
or magnetization contributes to increase the conductivity.
The numerical differences between Eqs.~\eqref{eq:NMR} and \eqref{eq:equilibrium2} can be traced back to the expansion of the modified density of states $D_{\mathbf{p},s}$ 
appearing in the denominator of Eq.~\eqref{eq:currentsemi} and the inclusion of the magnetic moment in the semiclassical calculation.
We note as well that, as with conventional magnetorresistance, we expect that the term Eq.~\eqref{eq:equilibrium} is also supplemented by a Fermi surface contribution due to the Lorentz force when the Fermi surface is anisotropic~\cite{PM10}.\\ 

We conclude this section by computing the center-of-mass velocity in this approach, an observable quantity in cold atomic experiments.
From Eq.~\eqref{eq:vcm}, to obtain the center-of-mass velocity and relate it to the conductivity we must compute the electron density $n$
that to leading order reads
\bea
\label{eq:density}
n=\sum_{s}\int \dfrac{d\mathbf{p}}{(2\pi)^3} D_{\mathbf{p},s}f^{s}_{\mathbf{p}}\sim \dfrac{\mu^3}{3\pi^2v^3\hbar^3}.
\eea
The relevant center-of-mass velocity component is obtained by inserting Eqs.~\eqref{eq:equilibrium} and \eqref{eq:density} into Eq.~\eqref{eq:vcm} to obtain 
\be
\label{eq:vcmz}
\delta v^{z}_{\mathrm{c.m.}}=\dfrac{\delta j_{z}}{n}=  -\tau\left( \dfrac{e^2v^2}{\mu}+\dfrac{2}{5}\dfrac{\hbar^2 v^6 e^4}{\mu^5} B_{5}^{2}\right)E
\ee
which is valid in the absence of magnetic field ($\mathbf{B}=0$). 
Therefore, through Eq.~\eqref{eq:vcmz} a cold atomic experiment can probe the anomalous longitudinal conductivity $\sigma\sim \mathbf{B}_{5}^2$ that we predict
by monitoring the center-of-mass motion of an atomic cloud.
%

%--------------------------------------------------------------
%
%------------------------------------------------------------
\section{Spectral properties of inhomogeneous semimetals \label{sec:spectral}}

\begin{figure*}%[b]
\includegraphics[width=\textwidth]{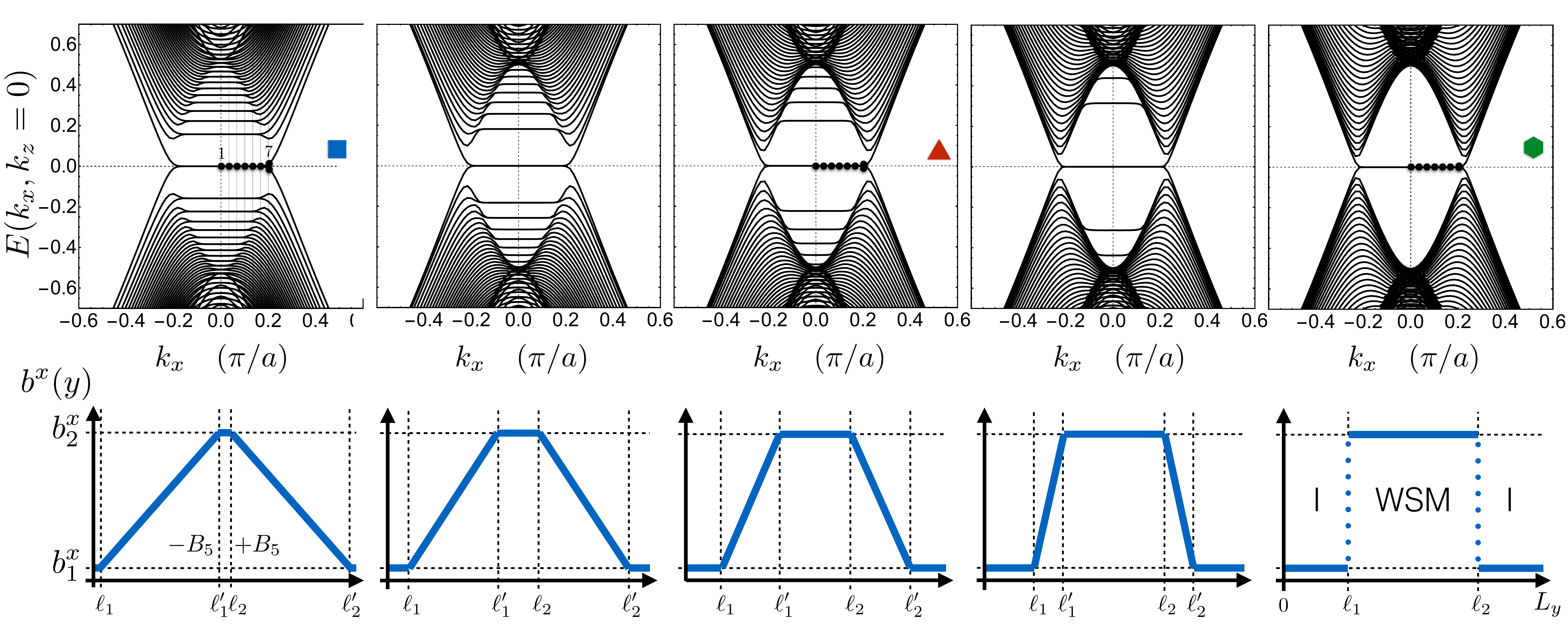}
\caption{The top row shows the energy spectra obtained by solving Eq. \eqref{eq:H4b} in the presence of a spatially varying $b^x = b^x(y)$, as a function of $k_x$ and for $k_z=0$. The schematic profile of $b^x(y)$ corresponding to each spectrum is presented below the energy panel. From left to right, the region of linear increase (decrease) of $b^x$, set by $\ell_1'-\ell_1$  ($\ell'_2-\ell_2$), gradually shrinks and becomes an abrupt stepwise increase (decrease) at $\ell_1$ ($\ell_2$). The latter case describes an interface between an insulator (I) and a Weyl semimetal (WSM) (all spectra are calculated for $m=0.5$), and the energy spectrum exhibits the zero energy Fermi arc surface states connecting the bulk nodes. The panel on the left shows the pseudo-Landau level structure of the low-energy states, arising as a result of a bulk axial magnetic field $B^z_5= \pm B_5$. As $\ell_1'-\ell_1=\ell_2'-\ell_2$ is decreased from left to right, the strength of the axial magnetic field is increased and the $n \ge 1$ pseudo-Landau levels (energy scales as $\sim \sqrt{n B_5}$) are pushed out of the spectrum. The $n=0$, however, remains at $E=0$ and morphs into the Fermi arc associated with the boundary surface shared by I and WSM. In these calculations $(b^x_1=0.0,b^x_2=1.0)$, $\ell_1'-\ell_1=\ell_2'-\ell_2 = (80,60,40,20)$, $(\ell_1,\ell_2)=(45,135)$ (right most panel), with system size $L_x \times L_y \times L_z = 120 \times 180 \times 120$ (all lengths measured in units of lattice constant $a$). The solid black dots in the upper panels indicate the (zero energy) states for which the wavefunction support is shown in Fig. \ref{fig:wavefunction}. 
}
 \label{fig:PLLa} 
\end{figure*}
%----------

Based on our field-theoretic and semiclassical discussion regarding the axial-gauge field coupling in Weyl semimetals, we now study the effects of magnetization or strain inhomogeneities within a microscopic lattice model realization. This allows us to numerically corroborate the arguments presented in the previous section. We will consider two canonical lattice models whose low-energy long-wavelength limit is given by Eq.~\eqref{eq:Weyleff}. For simplicity, we choose lattice realizations of Weyl semimetals with the minimum number of low-energy Weyl fermions, i.e., one Weyl fermion of each chirality, which implies that time-reversal symmetry is broken in both models. Our results, however, generalize to time-reversal symmetric (but inversion symmetry-breaking) models, and occasionally we will explicitly comment on such generalizations. 

The first minimal lattice model we use to describe Weyl fermions coupled to axial-gauge fields, is motivated by the solid state realizations of Weyl and Dirac semimetals. It can be obtained starting from a lattice-regularized version of a topological insulator $\bk \cdot \bp$ Hamiltonian~\cite{QHZ08}, such as Bi$_2$Se$_3$ or related materials. A simple generalization of this model has been used to describe the Dirac semimetals Cd$_2$As$_3$ and Na$_3$Bi~\cite{Vazifeh:2013fe,VGB15}. The model has four bands, originating from an orbital degree of freedom $A,B$ and spin $\up,\down$, and the corresponding electron operators are defined as $c_{\br} = (c_{\br A\up},c_{\br A\down},c_{\br B\up},c_{\br B\down})^T$. The Hamiltonian $H_{\text{4b}} $ of this four band model is the sum of two terms and is given by 
\be
H_{\text{4b}} = \sum_{\bk,j}  D^j(\bk) c^\dagger_\bk  \Gamma^j c_{\bk} + \sum_{\br,j} b^j(\br) c^\dagger_\br \Gamma_b \Gamma^j c_{\br}. \label{eq:H4b}
\ee
The first term is the topological insulator Hamiltonian with $\Gamma$-matrices $\Gamma^j = (\sigma^{z}s^{y},\sigma^{z}s^{x},\sigma^{y} s^0,\sigma^{x} s^{0}) $, $\Gamma_b = \sigma^y s^z$, and with the components of the $\mathbf{D}$ vector given by
\be
D^j(\bk)=-( \sin  k_xa, \sin  k_ya,\sin k_za , \sum_{i}\cos k_ia- M). \label{eq:Dvec}
\ee
Here we have set the kinetic energy scale $t=1$. The matrices $\sigma^{x,y,z}$ and $s^{x,y,z}$ are Pauli matrices acting on orbital and spin degrees of freedom, respectively ($\sigma^{0}$ and $s^0$ are identity matrices). The Dirac mass $M$, which describes a hybridization of the orbitals into bonding and anti-bonding states, controls whether the material is on a trivial- or topological-insulator phase (which may be strong or weak). Here, we set $M\equiv 3 +m $, such that $m=0$ corresponds to a Dirac semimetal state with a 3D Dirac point at $\bk=0$, and $m>1$ ($m<0$) corresponds to a trivial (topological) insulator. 

Whereas the first term in Eq. \eqref{eq:H4b} respects both time-reversal symmetry ($\mathcal{T}$) and inversion symmetry ($\mathcal{I}$), a nonzero $\bb = (b^x,b^y,b^z)$ breaks $\mathcal{T}$. To see how it is responsible for generating the Weyl semimetal phase, one may expand \eqref{eq:H4b} to linear order in $\delta \bk$ around $\bk=0$ and obtain
\be
H(\bk)=\sum_{j} (\delta k_j + b^j\Gamma_b) \Gamma^j + m\Gamma^4. \label{eq:H4bexpand}
\ee
This result shows that $\bb$ is responsible for the separation of the two nodes in momentum space, and thus couples as an axial gauge field. Indeed, Eq. \eqref{eq:H4bexpand} should be compared to the Weyl Hamiltonian of Eq.~\eqref{eq:Weyleff}, where the node separation was identified with the axial gauge field. In Eq. \eqref{eq:H4bexpand}, the matrix $\Gamma_b$ takes the role of giving the axial gauge field $\bb$ a different sign at the two nodes. It should be noted, however, that in Eq. \eqref{eq:H4bexpand} the node separation depends on both $\bb$ and $m$. In particular, the node separation is proportional to $\sqrt{|\bb|^2-m^2}$ and vanishes (i.e., the system is insulating) for $m > |\bb |$~\cite{Grushin:2012cb,VGB15}.

In the following, we will study Hamiltonian \eqref{eq:H4b} numerically in the presence of a spatially non-uniform $\bb(\br)$ which gives rise to axial gauge fields $\bB_5 \simeq \boldsymbol{\nabla}\times \bb(\br) \neq 0 $. It is important to note that $\boldsymbol{\nabla}\times \bb(\br)$ can only have the meaning of an axial magnetic field coupled to Weyl fermions in regions where $m < |\bb(\br) |$.  We always take $m\ge 0$, since we are not interested in the regime where $\bD$ describes a topological insulator. A full phase diagram of the model defined by Eq.~\eqref{eq:H4b} was described in Ref.~\onlinecite{VGB15}. 

The second lattice model we use to verify our results consists of effectively spinless electrons, has only two bands, and can be regarded as one of the two time reversal partners that compose the model in Ref.~\cite{DKLKSB2015}. It is therefore motivated by two practical considerations: {\it(i)} it is plausible to implement it in the cold atomic context, and {\it(ii)} it falls into the class of models tailored for the methods presented in Ref.~\cite{GJM16} to generate domain walls in topological systems. In this case, the electron operators are given by $c_{\br} = (c_{\br A},c_{\br B})$, where $A,B$ represent some generalized orbital degree of freedom, and the Hamiltonian takes the form
\be
H_{\text{2b}} = \sum_{\bk,j}  d^j(\bk) c^\dagger_\bk  \sigma^j c_{\bk}. \label{eq:H2b}
\ee
Here, $\sigma^z = \pm 1$ again represents the orbital degree of freedom, and the components of $\mathbf{d}$ are given by
\be
d^j(\bk)=-( \sin k_xa, \sin  k_ya,\sum_{i}\cos k_ia- M). \label{eq:dvec}
\ee
This model breaks time-reversal symmetry; it has a pair of linearly dispersing Weyl cones at $\mathbf{k} =\{0,0,\pm\cos^{-1}(M/t-2)\}$ for $1<\vert M/t\vert<3$,  two pairs of Weyl cones for $\vert M/t\vert<1$ and is a gapped insulator when $\vert M/t\vert>3$. The parameter $M/t$ controls the distance between Weyl nodes; interpolating between $M/J=2$ and $ M/t>3$  simulates the boundary between a Weyl semimetal with two nodes and an insulator. Therefore, we promote $M\to M(y)$ which sets the Weyl node separation as $\mathbf{b}(y)$ by
\be
\mathbf{b}(y)=(0,0,2\arccos(-2+M(y)/t)).
\ee 
We note that, although this model separates the Weyl nodes in $k_{z}$, any other separation direction can be chosen by redefining $\mathbf{d}$ appropriately.\\

From our calculations we find that both models qualitatively exhibit the same behaviour, so we will focus mainly on results obtained for the four-band model of Eq. \eqref{eq:H4b}. 

\begin{figure*}%[b]
\includegraphics[width=0.7\textwidth]{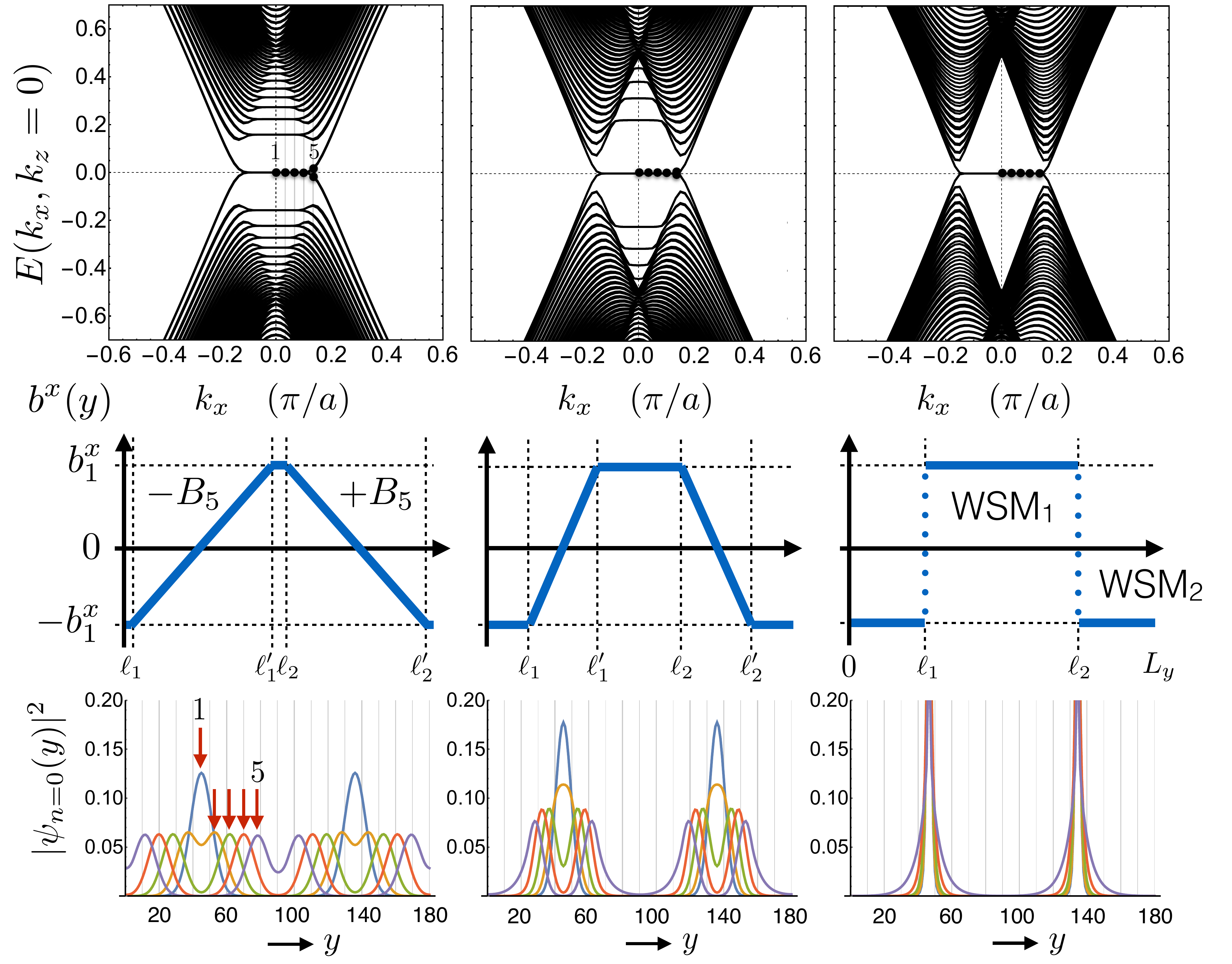}
\caption{Energy spectra (top row) obtained from Eq. \eqref{eq:H4b} in the presence of a spatially varying $b^x = b^x(y)$, shown in the corresponding central panels. The spectra were calculated using $m=0.0$ and $b^x_1 = 0.5$, such that the right most panels describe an interface between two Weyl semimetals (WSM$_1$ and WSM$_2$) with an inverted $\bb$ axial vector. As a result, at each of the two boundaries, $\ell_1$ and $\ell_2$, there will be two Fermi arcs, and the degeneracy of the zero energy states connecting the bulk nodes is doubled compared to Fig. \ref{fig:PLLa}. The energy level structure of the left most panel can be understood from the perspective of bulk axial magnetic fields. In each of the two regions, $B^z_5=+B_5$ and $B^z_5=-B_5$, two $n=0$ pseudo-Landau levels are present, one for each Weyl node, giving a total of four $n=0$ pseudo-Landau levels. The bottom panels show the support of the wavefunction of some $E=0$ states labeled by solid black circles as a function of $y$. The wave function support demonstrates that the Fermi arc states (on the right) are the Landau orbitals of the $n=0$ pseudo-Landau level (on the left). We used $\ell_1'-\ell_1=\ell_2'-\ell_2 = (80,40)$ for the first two columns, $(\ell_1,\ell_2)=(45,135)$ for the right most column, and a system size of $L_x \times L_y \times L_z = 120 \times 180 \times 120$.
}
 \label{fig:PLLb}
\end{figure*}

\subsection{Lattice pseudo-Landau Level structure of $\mathbf{B}_5$}

We begin by specifying the spatial profiles of $\bb(\br)$ that generate the axial magnetic configurations we will study. In what follows, we will always take $\bb(\br) = b^x(y)\hat{x}$, such that it only depends on $y$ and corresponds to an axial magnetic field along $z$. Since nonzero $b^x$ implies a separation of Weyl nodes along the $x$ axis in momentum space, $b^x(y)$ describes Weyl nodes whose separation $\Delta k_x$ depends on the $y$ coordinate. Note that this effectively corresponds to a Landau gauge, and consequently, $(k_x,k_z)$ remain good quantum numbers. Furthermore, it follows that the axial magnetic field $\diff \times \bb$ is orthogonal to the direction of node separation. The extension of the system in the $y$ direction is $L_y$ (similarly for $x$ and $z$), which we measure in units of the lattice constant $a$, and we assume $b^x(1)=b^x(L_y)$. 

We first consider the series of profiles $b^x(y)$ shown schematically in the bottom row of Fig. \ref{fig:PLLa}. The form of $b^x(y)$ is such that it increases linearly from $b^x_1$ to $b^x_2$ between $y=\ell_1$ and $y=\ell'_1$, then stays flat at $b^x_2$, and subsequently decreases linearly again from $b^x_2$ back to $b^x_1$ between $y=\ell_2$ and $y=\ell'_2$. As a result, between $\ell_1$ and $\ell'_1$ we have a constant and negative $\diff \times \bb \simeq \bB_5$, whereas between $\ell_2$ and $\ell'_2$ we have a positive constant $\diff \times \bb \simeq \bB_5$, with a magnitude that depends on the ratio $\Delta b^x/ \Delta \ell_{1,2}$, with $\Delta b^x= b^x_2-b^x_1$ and $\Delta \ell_{1,2} = \ell'_{1,2}- \ell_{1,2}$. For fixed $\Delta b^x$, the strength of the effective axial magnetic field increases for decreasing $\Delta \ell_{1,2}$. By taking $\Delta \ell_{1,2} \to 0$, as shown in the right most configuration of Fig. \ref{fig:PLLa}, we create two sharp interfaces between regions of constant $\bb = b^x_1\hat{x}$ and $\bb = b^x_2 \hat{x}$. We choose $m$ [see Eqs. \eqref{eq:H4b} and \eqref{eq:H4bexpand}] such that the latter corresponds to a boundary between a Weyl semimetal and an insulator, i.e., $m> b^x_1$ \cite{VGB15}. Thus, the configurations of Fig. \ref{fig:PLLa} interpolate between axial magnetic fields in the 3D bulk of the system, and a sharp 2D boundary between a Weyl semimetal and an insulator. The axial magnetic field is gradually confined to a 2D surface while increasing its strength. The energy spectra corresponding to these axial magnetic field profiles are shown in the top panels of Fig. \ref{fig:PLLa}, where we have set $b^x_1=0.0$, $b^x_2=1.0$, and $m=0.5$.

Before turning to an analysis of the energy spectra obtained from the lattice model, it is useful to recall the Landau level spectrum associated with axial magnetic fields in the continuum, cf. Eq. \eqref{eq:Weyleff}. We will refer to these Landau levels as pseudo-Landau levels. For Weyl fermions coupled to a uniform axial magnetic field $\bB_5$ along the $z$ direction, $B^z_5$, the $n\ge 1$ pseudo-Landau levels have energies $E_{n\pm}(k_z)=\pm\hbar v\sqrt{k_z^2+2|B^{z}_{5}|n}$ and thus disperse in $k_z$. In addition, there is an $n=0$ or zeroth pseudo-Landau level with energy $E_0(k_z)= \text{sgn}(B^z_5)\hbar v k_z$. Importantly, all energy levels are doubly degenerate: one for each of the two Weyl nodes. In particular, the two Weyl nodes, which have opposite chiralities, have \emph{the same} $E_0(k_z)$ dispersion of the zeroth Landau level. In contrast, in the presence of a (vector) magnetic field $B^z$, the chirality of the $n=0$ branch, i.e., the upward or downward slope as a function of $k_z$, depends on the chirality of the Weyl nodes~\cite{NielNino83}.  

The pseudo-Landau level structure of the continuum is reflected in the low-energy part of the lattice energy spectra shown in Fig. \ref{fig:PLLa}. In particular, in the left most panel we observe both a flat branch of zero energy states and flat branches of states at higher energies. These can be identified with pseudo-Landau levels since these do not disperse in $k_x$. Moreover, labeling the flat branches of states at nonzero energies by an index $n$, we find that the energies indeed scale as $\sim \sqrt{n}$. In the right most panel, where the axial magnetic field is confined to the sharp boundary between the Weyl semimetal and the insulator, we still find a branch of zero energy states. The flat branches at higher energies are absent. The zero energy states are simply the Fermi arc surface states, which must exist at such a surface boundary due to the topology of the Weyl nodes and they connect the projections of the bulk Weyl nodes. Figure \ref{fig:PLLa} thus suggests that as $\ell'_{1,2}- \ell_{1,2}$ is gradually taken to zero, confining the axial magnetic field to a narrower region along $y$ while increasing its magnitude, the $n=0$ pseudo-Landau level becomes the Fermi arc. The energy of the higher $n \ge 0$ pseudo-Landau levels scales as $\sim \sqrt{B_5}$, and are therefore ``pushed'' out of the spectrum as $B_5$ increases, as can be observed from left to right in Fig. \ref{fig:PLLa} .

The key implication of Fig. \ref{fig:PLLa} is that the Fermi arc surface states of a Weyl semimetal can be thought of as an $n=0$ pseudo-Landau level corresponding to an axial magnetic field spatially confined to the surface boundary, i.e., $\sim B^z_5\delta(y-\ell_{1,2})$. The chirality of the Fermi arc states, i.e., their dispersion in $k_z$ (discussed in more detail below), corresponds to the chirality of the $n=0$ pseudo-Landau level modes and depends on $\text{sgn}(B^z_5)$.

\subsection{Fermi arcs as $n=0$ pseudo-Landau Level}

To study this correspondence in more detail, we now consider a different set of $b^x(y)$ profiles, which are shown in the middle row of Fig. \ref{fig:PLLb}. As is schematically demonstrated, in this set of profiles $b^x(y)$ increases and decreases linearly from $-b^x$ to $b^x$, and we take $b^x=0.5$. Furthermore, $m$ is set to zero, such that the profiles interpolate between bulk axial magnetic fields on the left, and an interface between two Weyl semimetals with inverted Weyl node separation $b^x$ on the right. The corresponding energy spectra are shown in the top panels of Fig. \ref{fig:PLLb}. 

Focusing first on the right panel, i.e., the sharp interface between two Weyl semimetals with inverted $b^x$, we find that the branch of zero energy states, which connect the projections of the bulk nodes, is fourfold degenerate. These are the Fermi arcs localized at the boundaries between the Weyl semimetals, two for each boundary. This is consistent with the number of arcs mandated by topology. Similar to Fig. \ref{fig:PLLa}, the spectrum in the left panel exhibits the pseudo-Landau level structure at low energies. The flat branch of zero energy states corresponding to the $n=0$ pseudo-Landau level is fourfold degenerate, in agreement with the degeneracy of the Fermi arcs ($2 \times 2 = 4$). The pseudo-Landau level degeneracy can be understood by recalling that each Weyl node contributes one Landau level, which is doubled due to the two spatially separated regions of $\pm B_5$. (The counting in Fig. \ref{fig:PLLa} is more subtle, which we explain below.)  From Fig. \ref{fig:PLLb} we again observe that the Fermi arcs are adiabatically connected to $n=0$ pseudo-Landau levels as the profile of $b^x(y)$ is varied. The $n \ge 1$ pseudo-Landau levels are pushed to higher energies (and are eventually absent from the spectrum) due to the increasing axial magnetic field strength.

More insight can be gained by studying the wavefunctions of the Fermi arc states and comparing them to Landau orbital wave functions. Recall that wavefunctions of the lowest $n=0$ Landau level orbitals are given by
\be
\Psi^{n=0}_{k_xk_z}(\br) \propto e^{ik_x x}e^{ik_z z} e^{-(y- k_x l_b^2)^2/2l_b^2}, \label{eq:landauwf}
\ee
where $l_b$ is the magnetic length. The Landau orbitals are centered around $y= k_x l_b^2$, which locks the $y$ coordinate to momentum $k_x$. The distance between the Landau orbitals along $y$, as well as the spread of the wavefunction, are determined by the magnetic length $l_b$. Since $l_b \sim 1/\sqrt{B}$ the magnetic length decreases as the field strength $B$ increases. 

In Fig. \ref{fig:PLLb} we plot the support of the wavefunctions of the zero energy states corresponding to different values of $k_x$. The states for which wave functions are shown are indicated by solid black dots in the upper panels. From the bottom left panel of Fig. \ref{fig:PLLb}, we see that at each momentum there are indeed four states, with a wave function support similar to that of Landau orbitals. At $k_x=0$, these are localized at the center of each of the two regions of positive and negative axial magnetic fields, $B^z_5= \pm B_5$. As $k_x$ increases, the support of the wave function is shifted along $y$, and in opposite directions for Landau orbitals associated with Weyl nodes of opposite chiralities. This result follows from the position-momentum locking expressed in Eq.~\eqref{eq:landauwf}. 

As the profile of $b^x(y)$ is changed from left to right, the axial magnetic field strength is increased and thus the effective magnetic length is decreased. This is clearly reflected in the wavefunction support of the zero energy states: the spread becomes narrower and they move closer together. Eventually, when the axial magnetic field is confined to the interfacial boundary between the Weyl semimetals at $\ell_1$ and $\ell_2$, the Landau orbitals are localized at the boundaries, and should be viewed as the wave functions of the Fermi arc states.

\begin{figure}%[b]
\includegraphics[width=0.7\columnwidth]{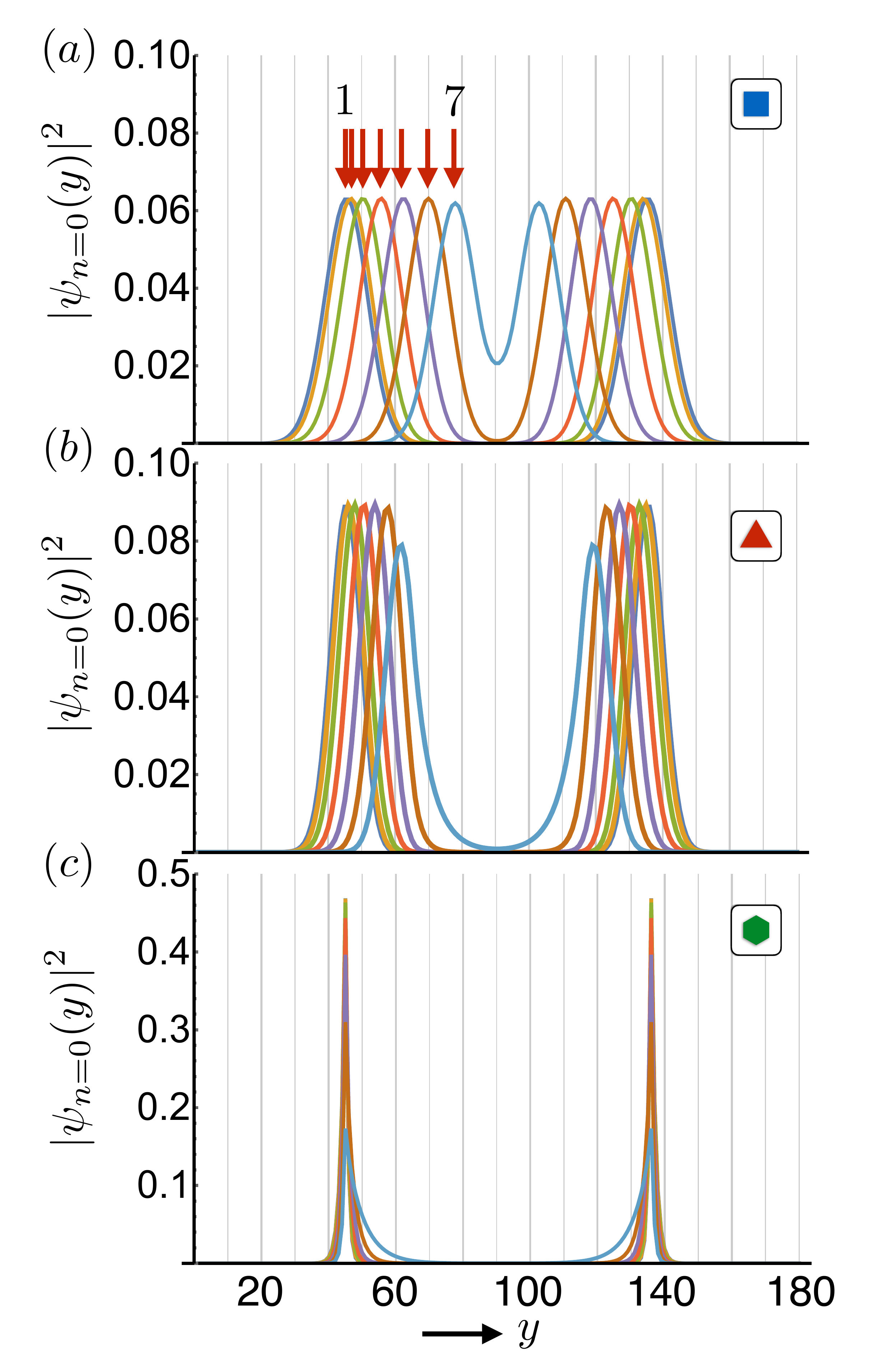}
\caption{Plot of the support of the wavefunction, as a function of $y$, of selected zero energy states corresponding to the spectra shown in Fig. \ref{fig:PLLa}. Panels (a), (b), and (c) correspond to left, middle, and right most panels of Fig. \ref{fig:PLLa}, respectively (graphically indicated by the square, triangle and hexagon). Here, in each panel, the seven different curves correspond to different values of $k_x$, which are indicated by solid black dots in Fig. \ref{fig:PLLa} and red arrows in (a). Two pseudo-Landau orbitals correspond to each value of $k_x$, as is most clearly seen in (a). This matches the degeneracy of $n=0$ pseudo-Landau level and is consistent with the number of Fermi arcs: one Fermi arc per boundary. Note that as compared to the bottom row of Fig.~\ref{fig:PLLb}, the Landau orbitals centered in the region where $b^x(y) < m $ are absent, which explains the difference in degeneracy (twofold vs. fourfold as discussed in the main text).
}
 \label{fig:wavefunction}
\end{figure}
%----------

An analogous picture arises when we plot the wavefunction support of zero energy states of the spectra in Fig. \ref{fig:PLLa}, which is presented in Fig.~\ref{fig:wavefunction}. Figure \ref{fig:wavefunction} (a) shows the wave functions of the $n=0$ pseudo-Landau level states of the left most panel in Fig.~\ref{fig:PLLa}, demonstrating the position-momentum locking $\langle y \rangle \propto k_x$. Note that there are only half the number of Landau orbitals as compared to Fig. \ref{fig:PLLb}, reflecting the different degeneracy of zero energy states: twofold versus fourfold. As is shown in Fig. \ref{fig:wavefunction}(a), the Landau orbitals in regions where $b^x(y)<m$ are absent. The region where $b^x(y)>m$ divides into a part where $B^z_5$ is positive and a part where $B^z_5$ is negative. In each region, both Weyl nodes contribute $n=0$ pseudo-Landau level orbitals, but of opposite momentum. Figure \ref{fig:wavefunction}(a) shows the contribution of only one Weyl node (chirality). 

Similar to Fig.~\ref{fig:PLLb}, we observe that in Fig.~\ref{fig:wavefunction}, as the magnetic length decreases towards zero from (a) to (c), the pseudo-Landau orbitals become the wave functions of the Fermi arc states localized at the surface boundary between the insulator and Weyl semimetal. From the perspective of Fermi arc surface states, the degeneracy of states is can be inferred from topology: the number of Fermi arcs at each boundary is equal to the Chern number change across the boundary. The Chern number can be defined when Weyl nodes of opposite (or different) Berry monopole charge are separated in momentum space. This suggests that, per the adiabatic connection between Fermi arcs and $n=0$ pseudo-Landau levels, the degeneracy of pseudo-Landau levels in the solid state systems has a topological origin.

\begin{figure}%[b]
\includegraphics[width=0.9\columnwidth]{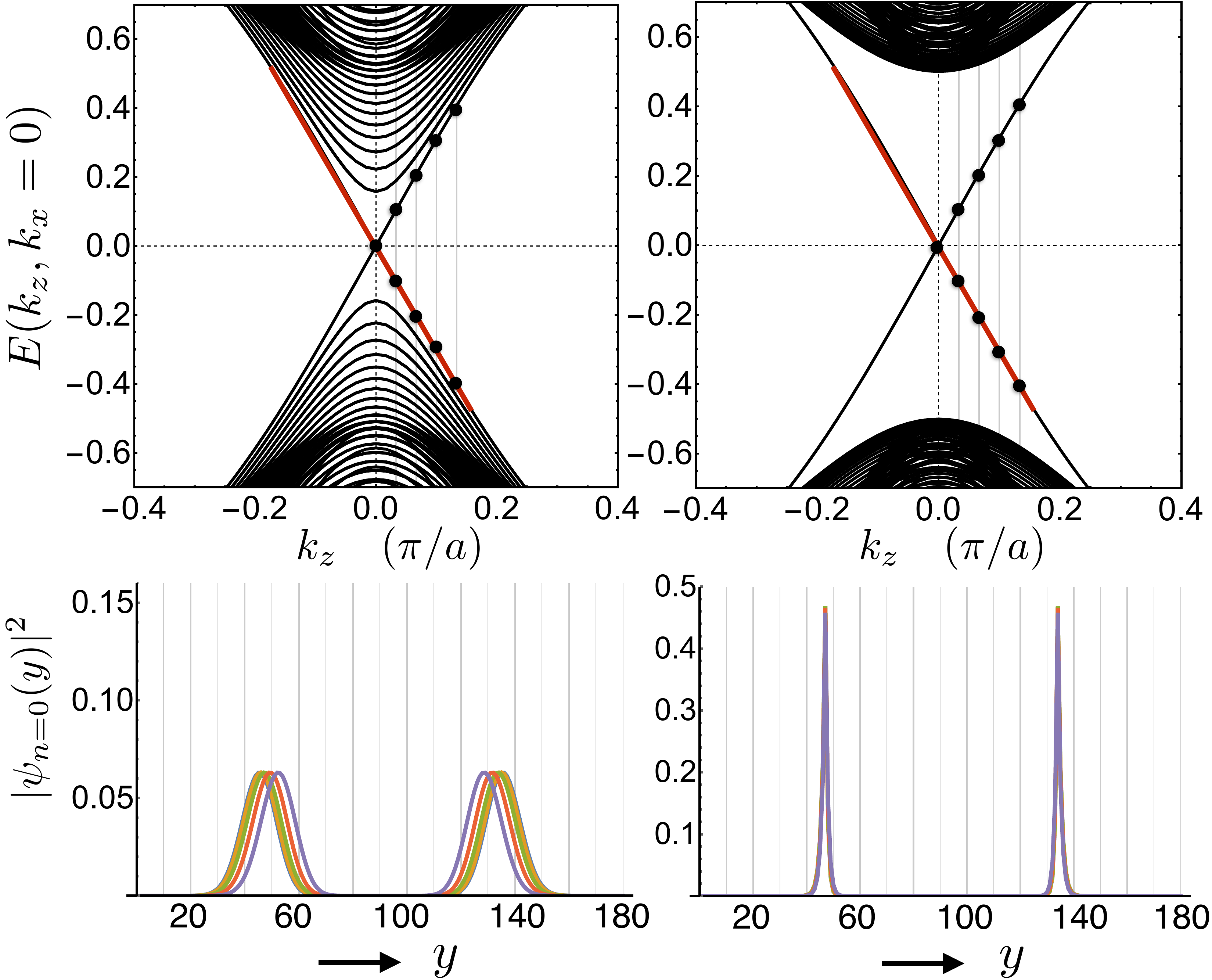}
\caption{Energy spectra, as a function of $k_z$ (with $k_x=0$), and wavefunction support obtained from Eq. \eqref{eq:H4b} in the presence of a bulk axial magnetic field $B_5^z$ (left) and for an interface boundary between an insulator and a Weyl semimetal (right). The spatial profile of $b^x(y)$ corresponding to the left and right panels here are given in the most left and most right bottom panels of Fig. \ref{fig:PLLa}, respectively. On the left, the low-energy branch of the spectrum has the structure of pseudo-Landau levels. The linearly dispersing chiral (black) and anti-chiral (red) modes are the $n=0$ pseudo-Landau levels. The wavefunction support of $n=0$ modes is shown for states with $k_z$ values indicated by solid black dots. On the right, the linearly dispersing modes correspond to the Fermi arc surface states and are sharply localized at the boundary interfaces $\ell_1$ and $\ell_2$ (see Fig. \ref{fig:PLLa}). The same systems sizes as in Figs. \ref{fig:PLLa} and \ref{fig:PLLb} are used.
}
 \label{fig:kz}
\end{figure}

The pseudo-Landau levels of the continuum are flat as a function of $k_x$, but disperse in $k_z$. Therefore, it is useful to compare the spectra of the lattice model as a function of $k_z$. The energy spectra, obtained for the two $b^x(y)$ profiles of the left most and right most panel of Fig.~\ref{fig:PLLa}, are shown in Fig. \ref{fig:kz}. The low-energy part of the spectrum in the left panel of Fig.~\ref{fig:kz} clearly exhibits the pseudo-Landau level structure, with $n \ge 1$ pseudo-Landau level energies dispersing in $k_z$ as $\sim \sqrt{k^2_z+nB_5}$. In addition, there is an $n=0$ chiral ($E \propto +k_z$) and anti-chiral ($E \propto -k_z$) mode, which are localized in different spatial regions corresponding to opposite $B^z_5$. This is confirmed by the wave function support of states with different $k_z$ in the bottom panel. Note that the momentum $k_z$ is not locked to the $y$ coordinate. The right panel of Fig. \ref{fig:kz} shows the dispersion of the Fermi arc states in $k_z$. The wave functions (bottom panel) are localized at the surface boundaries. Once more, one may think of the Fermi arcs as the $n=0$ chiral (and anti-chiral) modes of the $n=0$ pseudo-Landau level. The difference in the spread of the wave functions between the bottom left and bottom right panels originates from the difference in effective magnetic length.

\begin{figure}%[b]
\includegraphics[width=\columnwidth]{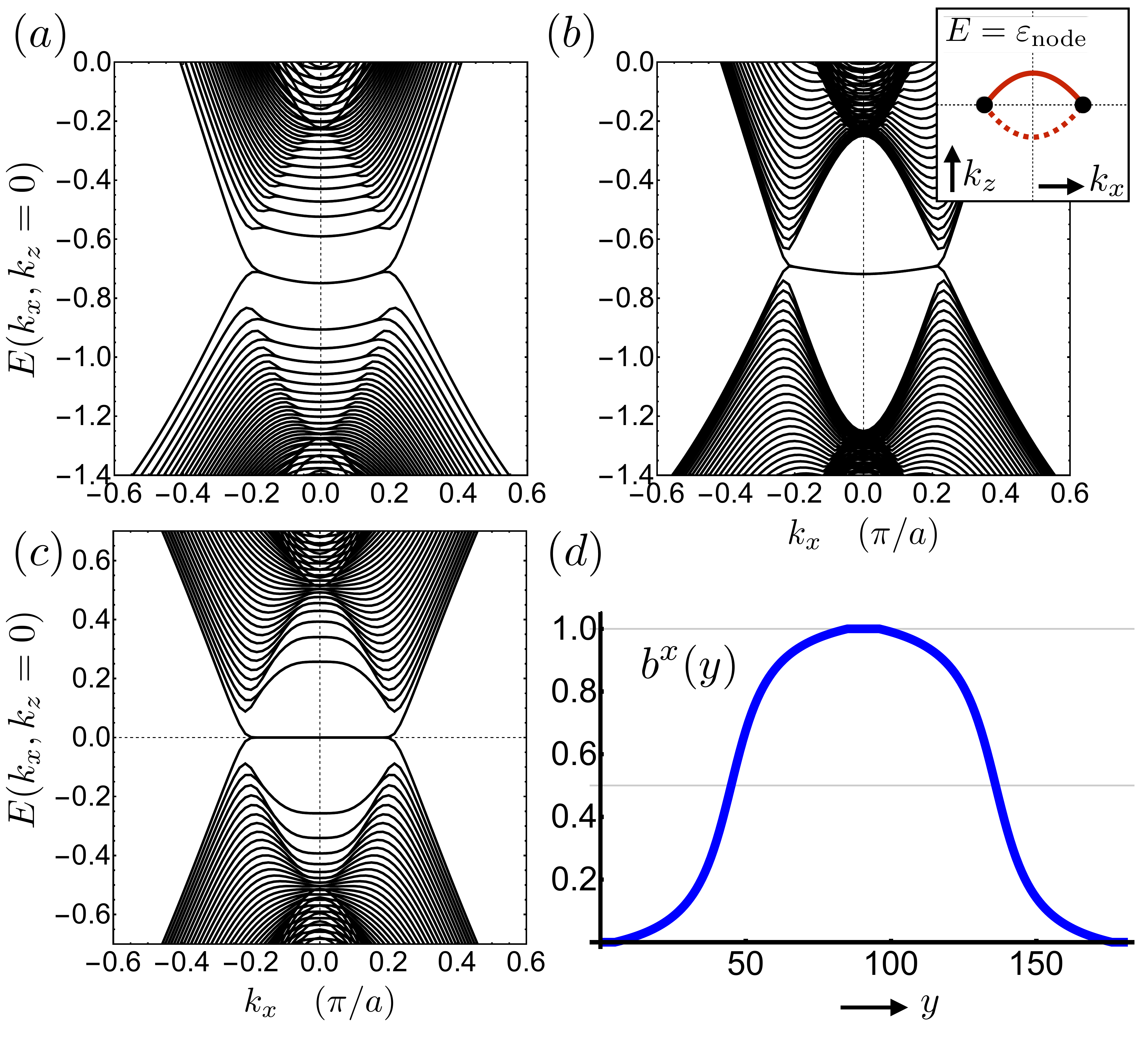}
\caption{Effect of the Fermi arc curvature and deviations from constant axial magnetic fields. (a,b) Plot of the energy spectra with $-\delta t \sum_i \cos k_ia \sigma^0s^0$ added to Eq. \eqref{eq:H4b} and $\delta t$ set to $\delta t = 0.25$. The $b^x(y)$ profiles corresponding to (a) and (b) are given in the most left and most right bottom panels of Fig. \ref{fig:PLLa}, respectively, with $m=0.5$. The effect of finite $\delta t$ is to give $k_x$ dispersion to the pseudo-Landau levels (a) and the Fermi arc states (b). The inset of (b) shows the Fermi arc states in the $k_x$-$k_z$ plane at the node energy $\varepsilon_{\mathrm{node}}$. (c) Energy spectrum obtained from Eq. \eqref{eq:H4b} with $b^x(y)$ as shown in (d), with $m=0.5$. Profile (d) leads to a non-uniform $\bB_5=\bB_5(y) \simeq \diff \times \bb  $, which is reflected in (c) where the $n\ge 1$ pseudo-Landau levels disperse in $k_x$ due to position-momentum locking. }
 \label{fig:smoothB5}
\end{figure}

At this stage, two remarks regarding the generality of our results are in order. The first concerns the profiles of $b^x(y)$ shown in Figs. \ref{fig:PLLa} and \ref{fig:PLLb}. In all these cases the increase or decrease of $b^x(y)$ is chosen to be linear, giving rise to a constant $B^z_5$. In solid state materials, axial vector potentials are due to strain or magnetization inhomogeneities, and one may expect the axial magnetic field strength to have a more general dependence on position, i.e., $\bB_5 = \bB_5(\br)$. To verify the application of our results to the more general case, we calculate the energy spectrum of Hamiltonian \eqref{eq:H4b} with $b^x(y)$ as shown in Fig. \ref{fig:smoothB5}(d). The corresponding spectrum is shown in Fig. \ref{fig:smoothB5}(c), which demonstrates two important features. First, the $n=0$ pseudo-Landau level remain non-dispersive in $k_x$ and at zero energy (for $k_z=0$). Second, the higher pseudo-Landau levels appear to have acquired a dispersion and shifted in energy. This is in agreement with the energies of the pseudo-Landau levels, $E_0 \propto \text{sgn}(B^z_5) k_z$ and $E_{n\pm}\propto \sqrt{k_z^2+2|B^{z}_{5}|n}$, respectively. Replacing $B^z_5$ by $B^z_5(y)$ and noting that $y \propto \text{sgn}(B^z_5) k_x$ due to Eq. \eqref{eq:landauwf}, it follows that the $n\ge 1$ pseudo-Landau levels should disperse in $k_x$, whereas the energy of the $n=0$ pseudo-Landau level should not change. This shows that our results hold for a general axial magnetic field configuration.

The second remark concerns the shape of the Fermi arcs. In Figs. \ref{fig:PLLa}, \ref{fig:PLLb}, and \ref{fig:kz}, Fermi arcs states at zero energy connect the bulk Weyl nodes in a straight line located on the $k_x$ axis. This is due to a non-fundamental symmetry of the model \eqref{eq:H4b}. In general, the shape of the Fermi arcs is not restricted and the condition $E(k_x,k_z)=\varepsilon_{\text{node}}$, where $\varepsilon_{\text{node}}$ is the energy of the bulk nodes (assuming they are both at the same energy), may define a curve connecting the bulk nodes with arbitrary shape. To study the more general case, we add  $ -\delta t \sum_i \cos k_ia \sigma^0s^0$ to Hamiltonian \eqref{eq:H4b} and calculate the energy spectra for two different $b^x(y)$ profiles. As shown in Figs. \ref{fig:smoothB5}(a) and (b), both the pseudo-Landau levels (a) and the Fermi arcs (b) disperse in $k_x$, implying that the Fermi arcs trace out a curve of the form shown in the inset of (b). We conclude from Figs. \ref{fig:smoothB5}(a) and (b) that the correspondence between Fermi arcs and $n=0$ pseudo-Landau levels holds for more general Fermi arc shapes.

\subsection{Calculation of the bound current density}

As pointed out in section~\ref{sec:transport}, the axial vector potential $\mathbf{b}$ in time-reversal broken Weyl semimetals physically corresponds to a magnetization vector, and is therefore expected to give rise to bound currents $\mathbf{j}_b$. 
Generally, the total current in an electronic system with a finite magnetization can be expressed as a sum of bound $\mathbf{j}_b$ and free current $\mathbf{j}_f$ densities~\cite{LANDAU1984xi}. The former has the property that it must average to zero over the system's volume, but it is allowed to be non-zero locally. 
It therefore can be expressed as the curl of a local vector $\mathbf{M}(\mathbf{r})$, the magnetization, such that $\mathbf{j}_b=\boldsymbol{\nabla}\times \mathbf{M}$. In the case where $\mathbf{M}$ is constant in the bulk, the bound currents only exist as surface currents, $\mathbf{j}^{\mathrm{surf}}_{b}$, that are perpendicular to the surface normal $\hat{\mathbf{n}}$ such that $\mathbf{j}^{\mathrm{surf}}_{b}=\hat{\mathbf{n}}\times\mathbf{M}$.

\begin{figure}%[b]
\includegraphics[width=\columnwidth]{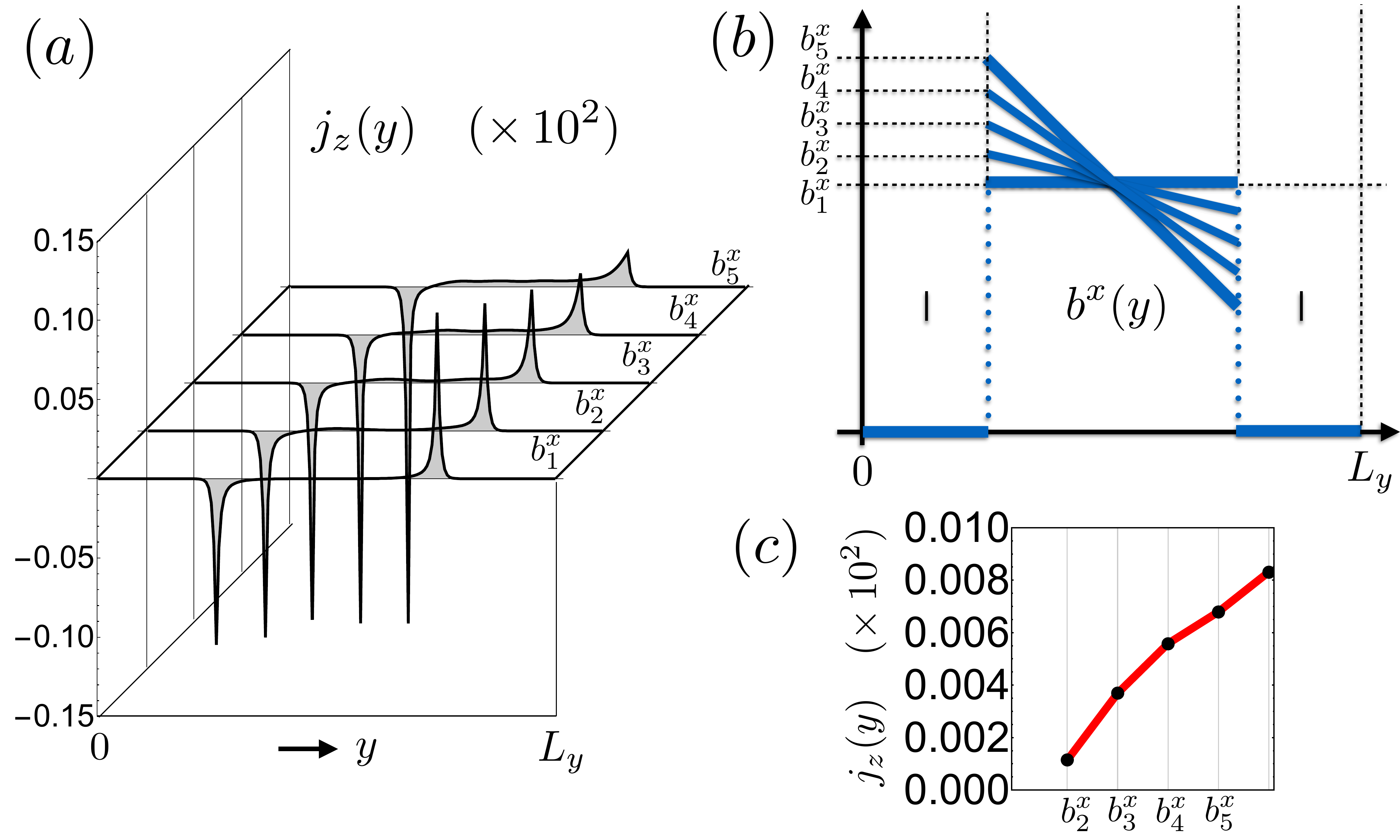}
\caption{Bound current density in an inhomogeneous Weyl semimetal: 
Panel (a) shows the five different current densities $j_z(y)$ corresponding to the five different profiles of
the axial vector potential shown in (b) labeled $b^{x}_a$ with $a=1,2,3,4,5$.
Larger values of $a$ support larger axial magnetic fields $\mathbf{B}_5=\boldsymbol\nabla \times \mathbf{b}$ which create 
bulk pseudo-Landau levels (see Fig.~\ref{fig:2weyls} (c)) that compensate for the missing current density at the boundary. 
Plot (c) shows the bound current density at a point $y=L_y/2$ belonging to the Weyl semimetal bulk for different values of $b^x_{a}$. 
}
 \label{fig:current}
\end{figure}

We now provide numerical evidence that justifies the interpretation of the second term in Eq.~\eqref{eq:NMR}, i.e., $\mu\mathbf{B}_5$, as a bound current density, by calculating the current density from our microscopic lattice models. 
To this end, we consider five different linear profiles of the axial vector potential of the form $\mathbf{b}=b_{a}^{x}(y)\hat{x}$, parametrized by $a=1,2,3,4,5$.
These trace out a finite Weyl semimetal with a Weyl node separation that increases linearly in the $y$ direction [see Fig.~\ref{fig:current} (b)]. 
The current density $j_z(y)$ is computed through the expression
\bea
\nonumber
j_z(y) &=& \langle \hat{J}_z(y) \rangle \nonumber \\
\nonumber
& = &\frac{1}{L_yL_z} \sum_{n,k_x,k_z}\left\langle u^n_{k_xk_z} \right| J_z(y)  \left | u^n_{k_xk_z} \right\rangle f(\varepsilon^{n}_{k_xk_z})\\
\label{eq:current}
\eea
where $J_z(y) = \partial H_{k_xk_z}(y)/\partial k_z$ is the current operator, $ \left | u^n_{k_xk_z} \right\rangle$ are the single particle eigenstates and $f(\varepsilon^{n}_{k_xk_z})$ is the Fermi-Dirac distribution function evaluated at the $n-$th eigenvalue $\varepsilon^{n}_{k_xk_z}$. 
Its real space distribution is shown in  Fig.~\ref{fig:current} (a) for different values of $b_{a}^x$, with an offset for clarity.
From Fig.~\ref{fig:current} (a), we observe two main features.
Firstly, for the flat profile $b_{1}^x$, the current is localized at the boundaries with equal weight but opposite sign.
This is entirely consistent with what is expected of a bound current; for a constant magnetization the bound currents are localized at the interface
and are normal to it ($\mathbf{j}^{\mathrm{surf}}_{b}$ explained above).
Secondly, as the slope increases (profiles $b_{a\neq 1}^x$), we observe that the weight associated with one boundary is transferred to the bulk, but the total current density remains zero overall.

\begin{figure}%[b]
\includegraphics[width=0.9\columnwidth]{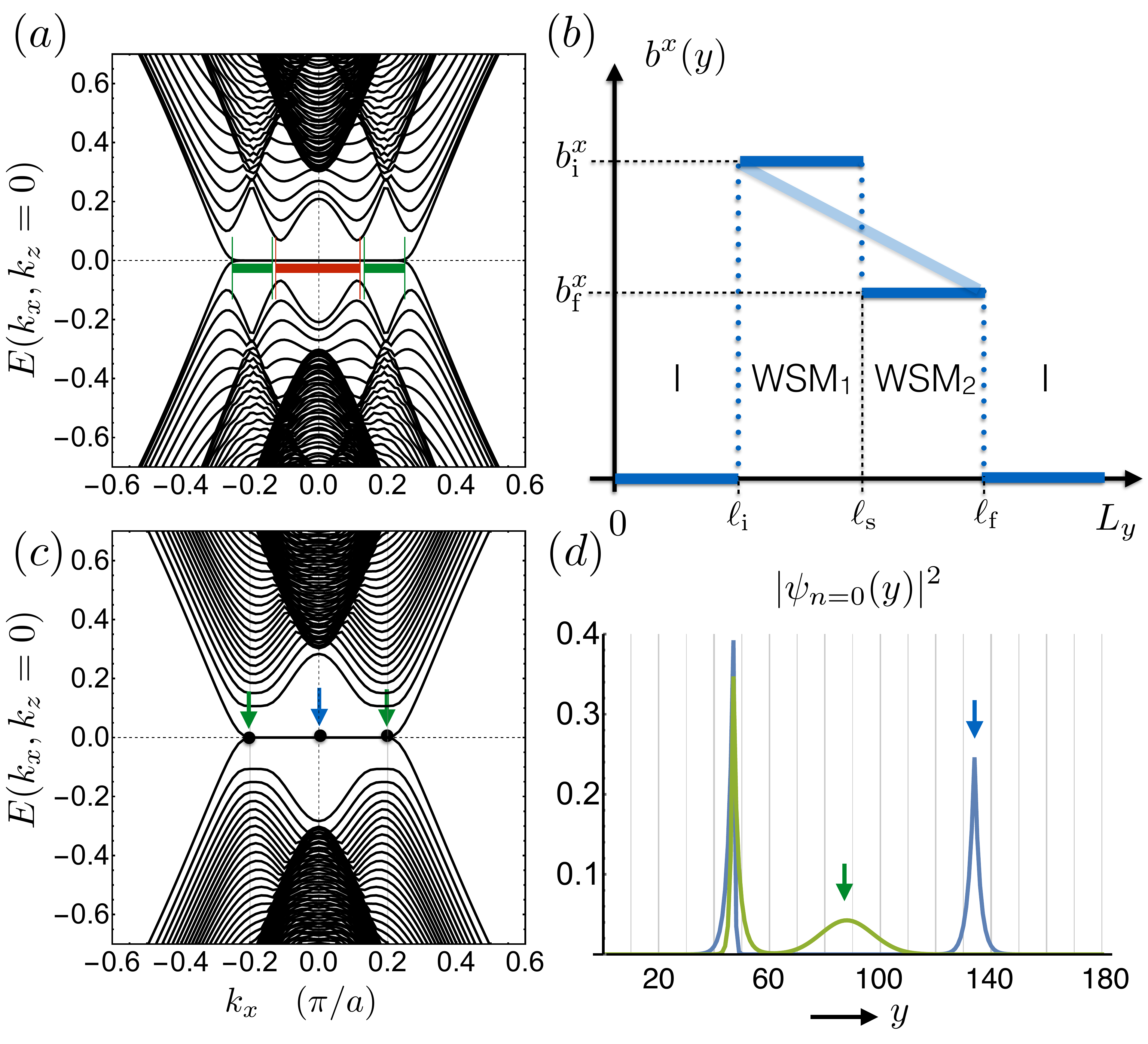}
\caption{Plot of the energy spectra (left panels) as a function of $k_x$ obtained for the $b^x(y)$ profiles shown in (b). The first case (a) corresponds to a heterostructure of insulator (I) and a Weyl semimetal in which the separation of Weyl nodes changes abruptly from $b^x_{\mathrm{i}}$ (WSM$_{1}$) to $b^x_{\mathrm{f}}$ (WSM$_{2}$) at $\ell_s$; see Eq. \eqref{eq:strainprof}. 
In (a), the zero energy states in the green region are associated
with the ``mini-arcs'' localized at $l_s$, whereas the states marked by
red are associated with the interface between the Weyl semimetals
and the insulator. In the second case (c) the separation of Weyl nodes is changed linearly between $\ell_\mathrm{i}$ and $\ell_\mathrm{f}$, see Eq. \eqref{eq:2weyls}, as shown by the dark blue line. The wave function support of states indicated by solid black dots in (c) is presented in (d). 
}
 \label{fig:2weyls}
\end{figure}

In order to understand both of these features in more detail, we express each of these linear profiles mathematically as
\begin{eqnarray}
\nonumber
b_{a}^{x}(y)=\left(\dfrac{b^x_{\mathrm{f}}-b^x_{\mathrm{i}}}{(\ell_\mathrm{f}-\ell_\mathrm{i})}(y-\ell_\mathrm{i})+b^x_{\mathrm{i}}\right)\left[\Theta(y-\ell_\mathrm{i})-\Theta(y-\ell_\mathrm{f}))\right].\\
\label{eq:strainprof}
\end{eqnarray}
which is schematically shown in Fig.~\ref{fig:2weyls} (b) by the light blue curve.
The corresponding $\mathbf{B}_{5}=\boldsymbol\nabla \times \mathbf{b}$ profile is set to a constant in the bulk (which is zero for the flat profile outlined by $b^x_{1}$ and non-zero for $b^x_{a\neq 1}$), and to two Dirac delta functions of opposite sign corresponding to each boundary.

Setting $b^x_{\mathrm{i}}=b^x_{\mathrm{f}}= b_{1}^x$ results in the spectral profile already discussed in Fig.~\ref{fig:PLLa} (lower row, right most panel): a Weyl semimetal with two Fermi arc surface states.
As chiral surface states, the two Fermi arcs carry two compensating current densities, consistent with what is observed for the $b_{1}^x$ case shown in Fig.~\ref{fig:current} (a). 
As the difference $b^x_{\mathrm{f}}-b^x_{\mathrm{i}}$ increases the profile traces higher values of $a$ in $b^x_{a\neq 1}$ and an increasing and constant $\mathbf{B_{5}} = \boldsymbol\nabla\times \mathbf{b}$ emerges in the bulk. 
The relevant energy spectrum for this situation is shown in Fig~\ref{fig:2weyls} (c); pseudo-Landau level emerge, and the 0th chiral pseudo-Landau level seems indistinguishable from a Fermi arc.

In fact, it is possible to distinguish the surface and bulk contribution due to position-momentum locking. As was discussed following Eq. \eqref{eq:landauwf}, the average guiding center position $\left\langle y \right\rangle$  is tied to the momentum $k_x$ 
such that $\left\langle y \right\rangle \sim k_x$.
Thus, by tracking the dependence of the wave-function of the zeroth Landau Level $\Psi_{n=0}(y)$ as a function
of $k_x$ we can extract real-space information. 

In Fig~\ref{fig:2weyls} (d) we show $|\Psi_{n=0}(y)|^2$ for two representative values of $k_x$ marked in Fig~\ref{fig:2weyls} (c). 
At $k_{x}=0$ the wave-function only has a finite weight at the edges, outlining a purely surface state. 
However, when $b^x_{\mathrm{f}}<|k_{x}|<b^x_{\mathrm{i}}$ the wave function acquires weight in the bulk associated with the 0th pseudo-Landau level emerging in this momentum space region. 
Importantly both regions are continuously connected as a function of $k_x$ and thus, from position-momentum locking, they are also connected in real space.
The bulk $\mathbf{B}_5$ creates bulk pseudo-Landau levels that connect to the surface arcs as a consequence of position-momentum locking.
This is a central result of this work and it allows us to understand the different instances of Fig.~\ref{fig:current} (a):
as $\mathbf{B}_5$ increases, a 0th bulk pseudo-Landau level forms that connects the unbalanced surface states.
The current density is sensitive to this fact; the current density lost by one Fermi arc is compensated and transferred to a bulk 0th pseudo-Landau level.
In other words the bound current at one surface is compensated by the sum of bound currents in the bulk and the remaining surface.

Taking one step back, our results can be summarized by the identification of the second term in Eq.~\eqref{eq:NMR} with a bound current due to the curl of a magnetization, $\mathbf{j}_b = \mu \mathbf{B}_{5} = \mathbf{\nabla}\times \mathbf{M}$. 
We have further corroborated this by explicitly checking that for small fields the magnitude of the bound current in the bulk of 
Fig.~\ref{fig:current} (a) increases linearly with $\mathbf{B}_{5}$ (cf. Fig.~\ref{fig:current} (c)) and it is only finite if $\mu\neq 0$.

All these arguments combined establish the three main points collected in the abstract:
{\it (i)} the Fermi arcs can be reinterpreted as $n=0$ pseudo-Landau levels resulting from a $\mathbf{B}_5$ confined to the surface 
{\it (ii)} a bulk $\mathbf{B}_5$ creates bulk-pseudo Landau levels that connect to the surface arcs as a consequence of position-momentum locking and
{\it (iii)} there are bound currents proportional to $\mathbf{B}_{5}$ and the chemical potential that average to zero over the sample, as occurs in magnetic materials. 

It is important to stress that the effects we discuss are sensitive to the pseudo-magnetic field direction, in particular whether it is parallel or perpendicular to the
Weyl node separation. In the parallel case, the bulk and surface behave as expected; the two bulk 0-th pseudo-Landau Levels have the same chirality, which is opposite to the chirality of the Fermi arcs at the boundaries. The surface spectral weight is evenly distributed between each arc, and compensates the bulk~\cite{PCF16}. In contrast, in the perpendicular case considered in this work, the surface states are allowed to have a chirality that matches that of the bulk 0-th pseudo-Landau levels, a fact that is enforced by position-momentum locking. The Fermi arcs at different surfaces do not have the same spectral weight, rendering the rich spectral structure reported here. 

\subsection{ Interface between two Weyl semimetals as a finite bulk $\mathbf{B}_5$}

Given the above discussion, it is interesting to consider the case of two Weyl semimetals of different Weyl node
separation that share a boundary. 
At this point, we can intuitively predict the outcome: if the system abruptly changes the Weyl node separation, this is equivalent to the appearance of a $\mathbf{B}_5$ that is confined to the interface.
This case occurs when the change from $b^x_{\mathrm{i}}$ to $b^x_{\mathrm{f}}$ is localized at a single point in space, as shown in Fig.~\ref{fig:2weyls} (b) dark curve.
It represents two Weyl semimetals with two constant axial vectors potentials $b^x_{\mathrm{i}}$ and $b^x_{\mathrm{f}}$ brought into contact. 
From the previous section, we expect that currents bound to the surfaces of the two semimetals have a different magnitude, 
and their difference is proportional to $\boldsymbol\nabla\times(\mathbf{b}_{\mathrm{i}}-\mathbf{b}_{\mathrm{f}})$. 
Alternatively, the extensions of their Fermi arcs in momentum space are different. Therefore, we expect that there is a finite amount of current bound to the interface, that compensates for that difference. 

From a spectral point of view, it is interesting to observe how the Fermi arcs behave in this case. 
To this end, we consider the profile of $b^{x}(y)$ shown by the dark curve in Fig.~\ref{fig:2weyls} (b). 
Mathematically it is described by
\bea
\nonumber
b^{x}(y)&=&b^x_{\mathrm{i}}\Theta(y-\ell_\mathrm{i})+(b^x_{\mathrm{f}}-b^x_{\mathrm{i}})\Theta(y-\ell_\mathrm{s})\\
&-&b^x_{\mathrm{f}}\Theta(y-\ell_\mathrm{f}).
\label{eq:2weyls}
\eea

The corresponding spectrum is shown in Fig.~\ref{fig:2weyls} (a) and is characterized by the appearance of``mini-arcs", fragments of longer arcs that are surface states belonging to the uniform system with the larger Weyl node separation $b_\mathrm{i}^x$. 
This situation is quite generic for an interface between two topological phases with the same topological invariant -  the surface states hybridize and gap out along the region of contact. 
For Weyl semimetals we can consider this from the perspective of assigning a Chern number to two-dimensional slices of momentum space that lie between two Weyl nodes~\cite{Wan2011}. The corresponding chiral edge states hybridize and gap out in the region of overlap in momentum space, which is equal to the length of the shorter axial vector potential $b_\mathrm{f}^x$. 

It is now straightforward to understand how the spectrum evolves when the interface between two Weyl semimetals is smeared across the entire sample, which corresponds to Fig.~\ref{fig:2weyls} (b) light blue curve. 
The spectrum in this case, presented in Fig.~\ref{fig:2weyls} (c), shows that the smeared interface
transforms the interfacial arc states at $l_s$ into zeroth pseudo-Landau Levels.
Moreover, this conclusion based on the spectral information of Fig.~\ref{fig:2weyls} (c) is corroborated by 
the wavefunction spread in real space shown in Fig.~\ref{fig:2weyls} (d). As discussed in the last section, the latter portrays how, as $k_x$ is changed,
the real space probability density shifts from edge to bulk.
We stress that this is another complementary instance of the discussion around Figs.~\ref{fig:PLLa} and~\ref{fig:PLLb}.
The surface Fermi arcs blend with the 0th-Landau levels within a single $b^{x}(y)$ profile, Fig.~\ref{fig:2weyls} (b) (light blue curve) rather than a sequence of them as in Figs.~\ref{fig:PLLa} and~\ref{fig:PLLb}.

Moreover, the comparison between the light blue and dark blue profiles in Fig.~\ref{fig:2weyls} (b) 
inspires the following physically appealing picture of how the current density profile is distributed for the linear profile studied in Fig.~\ref{fig:current} (b).
The region between $\ell_\mathrm{i}$ and $\ell_\mathrm{f}$ in Fig.~\ref{fig:2weyls} (b) (light blue curve) can be approximated by a collection of infinitesimal discontinuities in the spirit of the trapezoidal method for curve integration. At each discontinuity there are Fermi arcs that meet and annihilate with a part of the arcs of the neighboring layer. These bulk arcs carry current density and thus act as a collection of bulk sheets that ``leak" current density into the bulk from the surface.

\section{Discussion and conclusion}\label{sec:discuss}

In this work we studied inhomogeneous Weyl and Dirac semimetals with a space dependent Weyl node separation. We have discussed how such a scenario can arise either from inhomogeneous strain or magnetization in existing solid state systems as well as cold-atomic setups. Underlying our results is an axial magnetic field $\mathbf{B}_5$ that couples to the electronic degrees of freedom with two main experimental consequences for both Dirac and Weyl semimetals, which we have addressed in depth.

The first experimental consequence is a drastic change of the spectral properties of topological semimetals.
Using two different lattice models we have established two novel spectral features attributed to the emergence of $\mathbf{B}_5$:
{\it (i)} Fermi arcs are secretly 0th pseudo-Landau levels due to a finite and large $\mathbf{B}_5$ at the boundary and
{\it (ii)} bulk pseudo-Landau levels form due to $\mathbf{B}_5$ and compensate the difference in density of states 
of inequivalent Fermi arcs at opposite boundaries via position-momentum locking. 
Point {\it (i)} is an inevitable consequence of the boundary supporting a finite $\mathbf{B}_5$ since the Weyl node separation must vanish in vacuum. This allows us to identify the existence of Fermi arcs with the emergence of a boundary 0th pseudo-Landau level due to $\mathbf{B}_5$, even in the absence of bulk inhomogeneities.
Such a correspondence provides a novel perspective on surface physics of Weyl and Dirac semimetals. 
Point {\it (ii)} is particularly relevant for spectral probes like ARPES or STM that are sensitive to a modification of the electronic spectrum~\cite{LBM10,GKW12,PGF13}. Remarkably, for the particular case of time-reversal broken realizations of inhomogeneous Weyl semimetals, the axial magnetic field results in an inhomogeneous distribution of bound currents throughout the sample, which exist in equilibrium and average to zero over the entire volume. 
The appearance and field dependence of these bound currents in the lattice realization that we study is also consistent with our semiclassical treatment of the response.

We envision two plausible routes to probe the effects resulting from the discussed bound currents.
Magnetic sensors such as scanning superconducting quantum interference devices (SQUIDs) are a natural way to probe local distributions of bound currents. Such probes were proven very successful at detecting magnetization modulations and inhomogeneous current distributions at interfaces such as LAO/STO heterostructures, and in two- and three- dimensional topological insulators (cf. Refs.~\cite{kalisky2013locally,nowack2013imaging,wang2015observation}). It is important to emphasize that non-local currents due to an inhomogeneous magnetization and the bound currents due to $\mathbf{B}_{5}$ are physically analogous phenomena, and thus it is very likely for the latter to be detectable. The difference between bound currents in an ordinary magnetized material and bound currents within a Weyl semimetal is that those in the latter emerge from a unique coupling between the Weyl fermions and the background magnetization as described throughout this paper. 

Another promising alternative are torque experiments, which have in fact already been conducted with Weyl semimetals~\cite{MPR15}.  The magnetic torque $\mathbf{\tau}=\mathbf{M}\times\mathbf{B}$ is a direct measure of the magnetic anisotropy of a crystal, and thus can reveal noncompensation of surface magnetic domains.\\
%

%
%---------
\begin{figure}%[b]
\includegraphics[scale=0.235]{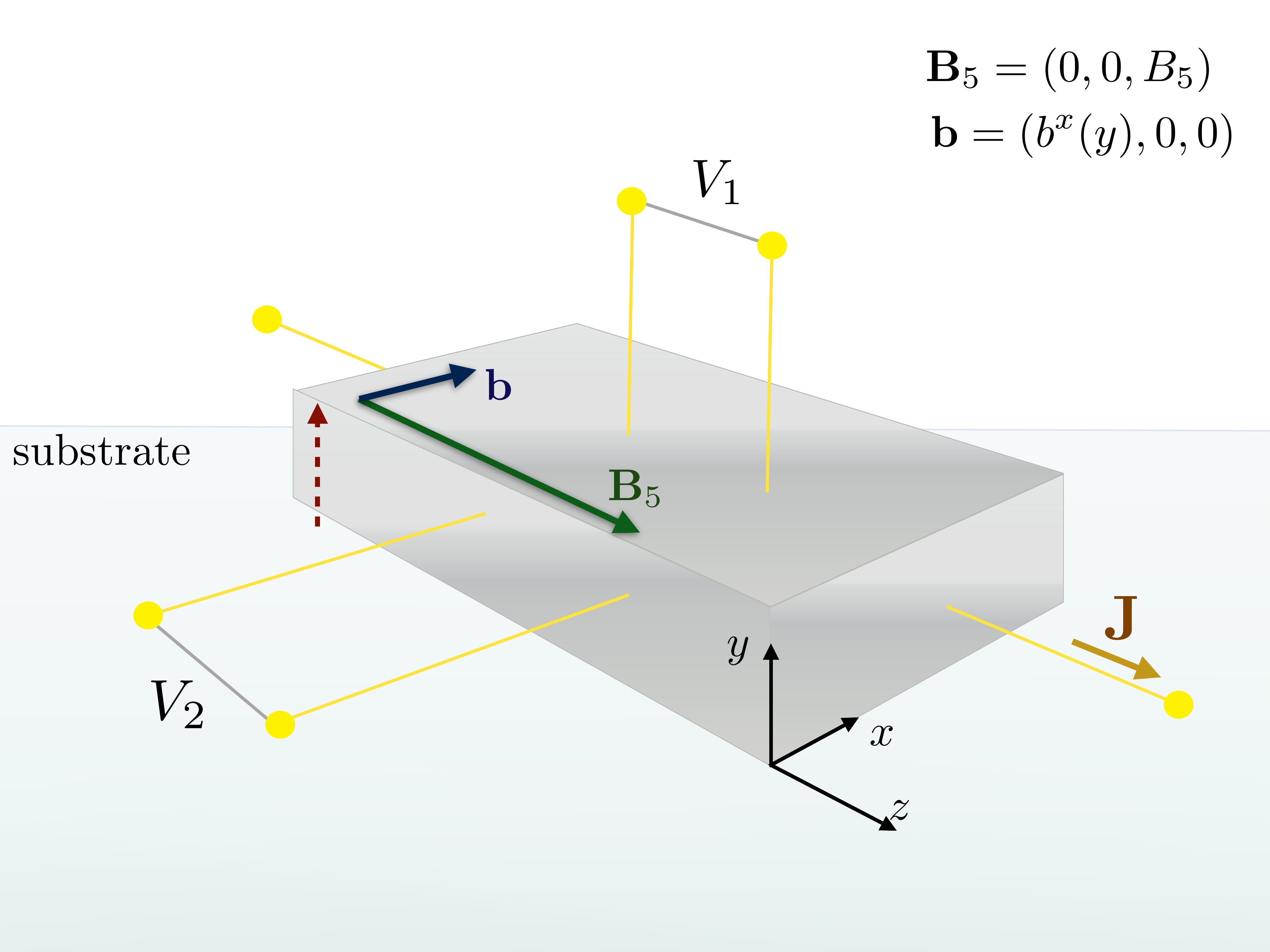}
\caption{A schematic point contact transport set-up to measure the current $\mathbf{J}$ (orange arrow) to determine the strain dependent conductivity $\sigma\sim \mathbf{B}_{5}^2$ predicted in this work. Strain is induced by the substrate with a small lattice mismatch that relaxes along the $y$ direction (red dashed arrow). If the Weyl node separation is $\mathbf{b}\parallel \hat{x}$ (blue arrow) then 
$\mathbf{B}_{5}\parallel \hat{z}$ (green arrow). Contacts $V_1$ ($V_{2}$) probe a surface with (without) Fermi arcs, resulting in an anisotropic contribution to the conductivity.}
 \label{fig:setup}
\end{figure}
%----------

Within our semiclassical treatment we have predicted a second experimentally relevant consequence, which is related to transport: inhomogeneities enhance the longitudinal conductivity in Dirac or Weyl semimetals as $\sigma\sim\mathbf{B}^2_{5}$. We find that this transport property is routed in a chiral pseudo magnetic effect similar to the enhancement of the magneto-conductance due to the existence of the chiral magnetic effect. We stress that this prediction is applicable to all kinds of topological semimetals. Since this enhancement is insensitive to the sign of $\mathbf{B}_5$, it will be generated by each pair of Weyl nodes, a number that is of order ten in realistic materials. 

A typical transport setup with point contact probes, schematically shown in Fig.~\ref{fig:setup}, suffices. It represents a thin film, or nanowire of a topological semimetal that is strained by a substrate with a small lattice mismatch. This setup can be particularly relevant for strained HgTe/CdTe heterostructures discussed in Ref.~\cite{RKY16}. 
As illustrated in Fig.~\ref{fig:setup}, point contact measurements have the additional freedom to choose the voltage probes to be on a surface with or without Fermi arcs ($V_{1,2}$ respectively), depending on wether the momentum space projection of the nodes on that particular surface coincides or not. Although for the TaAs class of materials all accessible surfaces host Fermi arcs, Dirac materials like Cd$_{3}$As$_2$ do present surfaces free of arcs. Such anisotropy highlights the difference between topological metals and more conventional strained semiconductors. While in the latter, conductance can also be enhanced via strain by modifying the band structure, changing the position of the voltage probes is not expected to result in anisotropic measurements if current jetting is negligible~\cite{SAW2015}.

The above considerations suggests a way to probe the surface bound currents. As we have discussed extensively in section \ref{sec:spectral}, surface currents on opposite surfaces in the presence of bulk strain will not compensate each other completely since their difference will be carried by the bulk pseudo-Landau levels. It could be expected that two distinct surfaces that are perpendicular to $\mathbf{B}_5$ will carry different currents. Put differently, the Fermi arcs of two opposite surfaces will not be of equal length in the presence of bulk strain. Thus, measuring an anisotropy in the surface currents could reveal the size of the bulk strain gradients. However, typical surface current signals are small, and thus this detection mechanism is less feasible in practice. Nonetheless we note that the special nature of the Fermi arcs can conspire to make the surface contribution sizable, as in other related situations~\cite{BHT16}. \\

In order to assess the significance of the effects we predict we now estimate the magnitude of a bulk axial-field $\mathbf{B}_{5}$. For a thin sample, it is expected that strain relaxes linearly with height over several unit cells if the lattice constant mismatch is small. For a typical sample, such as a Cd$_3$As$_2$ thin film, we can consider a height of the order of $\Delta L \sim 10$ nm,  and a conservative value of strain is translated into a change in lattice constant of $\sim 1\%$ but can be as large as $\sim10\%$. If we assume that the Weyl node separation spans typically a tenth of the Brillouin zone $|\mathbf{b}|\sim1/10$  \AA$^{-1}$, then the effective magnetic field is $|\mathbf{B}_{5}|\sim  \frac{\hbar\Delta b}{e\Delta L}\sim \frac{\hbar}{e}10^{15} $ m$^{-2}\sim 4$ T. Remarkably, this conservative estimate results in sizable magnetic fields, certainly above the detectable threshold of magnetic loops and SQUIDS.
These fields will also induce detectable changes in conductance that can be probed by growing samples with different $\Delta L$ that result in different intrinsic $\mathbf{B}_{5}$.

Cold atomic systems offer a controlled alternative. For these systems, we have proposed that the conductivity enhancement can be measured by monitoring the center-of-mass velocity $\mathbf{v}_{c.m.}$. This quantity is the quotient between the current density $\mathbf{j}$ and the particle density $n$, which will depend on the external fields in general. For small fields the density reduces to a trivial constant and the center-of-mass position is determined solely by the current density times a constant factor. This experiment could be performed in a cold atomic realization of the two band model used in this work. Moreover, this model, along with the family of two band models that host Weyl fermions~\cite{DKLKSB2015}, are ideal to apply the method proposed in Ref.~\cite{GJM16} to study inhomogeneities. It relies on implementing a space dependent offset between neighboring sites. Applying such an offset profile, which is feasible experimentally, results in a space dependent onsite potential ($M(y)$ in our two band model) leading to a finite axial magnetic field, as described in this work. These two considerations render cold atomic systems as a natural platform to engineer and observe the effects presented.\\

To conclude, we have shown that inhomogeneous strain and magnetization have profound and observable implications on the electronic spectrum and transport properties of Dirac and Weyl semimetals. Moreover, our general analysis provides an alternative angle to explore distinct topological to trivial and topological to topological interfaces, both smooth and abrupt~\cite{VGB15}. The fundamental correspondence between Fermi arcs and pseudo-Landau levels and their connection to bound currents departs from naive expectations and can inspire future theoretical and experimental studies of surface effects in Weyl and Dirac semimetals. \\

\emph{Note added:} While preparing this manuscript, we became aware of the recent related work by Pikulin \textit{et al.} \cite{PCF16} discussing the effect of strain generated by torsion in inversion breaking Weyl semimetal nanowires. Their findings are consistent with and complementary to our results. In particular torsion in wires generates an axial magnetic field $\mathbf{B}_{5}$ that is parallel to the Weyl node separation rather than the perpendicular component described here.

\section{Acknowledgements}

We are grateful to A. Cortijo and D. Pesin for their insightful feedback and for sharing results prior to publication. We also thank J. Analytis, Y. Ferreiros, N. Goldman, F. de Juan, K. Landsteiner, T. Neupert, D.~I.~Pikulin, A. C. Potter, J. E. Moore, T. Morimoto, M. A. H. Vozmediano, L. Wu, and S. Zhong for enlightening discussions. We thank Jens H. Bardarson for a related collaboration (Ref.~\cite{VGB15}) and discussions that inspired part of this study. A. G. G. acknowledges funding from the European Commission under the Marie Curie Programme Contract No. 653846. J.V. was supported by the Netherlands Organization for Scientific Research (NWO) through a Rubicon grant. R.I. acknowledges support from AFOSR MURI. A.V. is supported by NSF DMR-141134 grant.

%\bibliography{WSM,dynamics,WSM_WP,database.bib,TaP-references.bib}

\end{document}